# Capturing Richer Information—On Establishing the Validity of an Interval-Valued Survey Response Mode

**Zack Ellerby**

University of Nottingham, Lab for Uncertainty in Data and Decision Making, School of Computer Science

**Christian Wagner**

University of Nottingham, Lab for Uncertainty in Data and Decision Making, School of Computer Science

**Stephen Broomell**

Carnegie Mellon University, Department of Social and Decision Sciences

**Author Note:**

Corresponding author: Zack Ellerby

Email: zack.ellerby@nottingham.ac.uk

Address:

School of Computer Science

University of Nottingham

Jubilee Campus

Wollaton Road

Nottingham, NG8 1BB

## Abstract:

Obtaining quantitative survey responses that are both accurate and informative is crucial to a wide range of fields. Traditional and ubiquitous response formats such as Likert and Visual Analogue Scales require condensation of responses into discrete point values—but sometimes a range of options may better represent the correct answer. In this paper, we propose an efficient interval-valued response mode, whereby responses are made by marking an ellipse along a continuous scale. We discuss its potential to capture and quantify valuable information that would be lost using conventional approaches, while preserving a high degree of response-efficiency. The information captured by the response interval may represent a possible response range—i.e., a conjunctive set, such as the real numbers between three and six. Alternatively, it may reflect uncertainty in respect to a distinct response—i.e., a disjunctive set, such as a confidence interval. We then report a validation study, utilizing our recently introduced open-source software (DECSYS) to explore how interval-valued survey responses reflect experimental manipulations of several factors hypothesised to influence interval width, across multiple contexts. Results consistently indicate that respondents used interval widths effectively, and subjective participant feedback was also positive. We present this as initial empirical evidence for the efficacy and value of interval-valued response capture. Interestingly, our results also provide insight into respondents' reasoning about the different aforementioned types of intervals—we replicate a tendency towards overconfidence for those representing epistemic uncertainty (i.e., disjunctive sets), but find intervals representing inherent range (i.e., conjunctive sets) to be well-calibrated.

**Keywords:** questionnaire, survey, interval, ellipse, uncertainty, vagueness, ambiguity, sets

# Introduction

The collection, analysis, utilization (and monetization) of data, obtained largely from people, has increased exponentially over the last decades. This growth reflects rapid progress in information and communications technologies, which have been successfully exploited to improve access to these data. Many areas of research are already seeing great practical benefits from technological advances—consider the growing prevalence of remotely administrable digital questionnaires (Behrend, Sharek, Meade & Wiebe, 2011; Gnambs & Kaspar, 2015; Krantz & Reips, 2017; Schmidt, 1997). Moreover, development and evaluation of methods and best practices for collecting and interpreting survey response data have long been important topics, subject to both extensive discussion and empirical research (cf. Converse & Presser, 1986; Fowler, 1995; Groves et al., 2011; Krosnick & Fabrigar, 1997; Meade & Craig, 2012; Payne, 1951; Saris & Gallhofer, 2014; Thurstone & Chave, 1929). However, in stark contrast with the rapid development in methods of reaching people to acquire data, the predominant modes of capturing, encoding, and quantifying responses to social, behavioural, and psychological surveys remain fundamentally those developed early in the 20th century—i.e., ordinal, 'Likert-type' and continuous 'Visual Analogue' scales (Likert, 1932; Freyd, 1923). We propose that it is time to pause, take stock, and ask whether technological advances permit improvements to the *type* of data that can be efficiently collected from survey respondents, as well as the means to access and analyse them.

Specifically, while a broad range of specific aspects of questionnaire design have already been investigated in great detail, this paper focuses on another, which we believe has received relatively little attention given its potential significance. This is the efficient handling of uncertainty or range that may be inherent in individual responses—which may arise from a wide variety of factors, including: lack of knowledge or information available to the respondent, inherent randomness, variability, or vagueness in the answer, and ambiguity or ill-definition in the question. This paper begins by discussing the potential value of capturing such information through interval-valued responses, which may represent either 'conjunctive' or 'disjunctive' sets (cf. Couso & Dubois, 2014; Dubois & Prade, 2012), before contextualising this in comparison with the capabilities of conventional methods. It then puts forward and empirically evaluates an ellipse-based response mode—which is designed to efficiently capture and quantify the uncertainty and range associated with each response in the form of an interval, increasing both fidelity and information content. This method thus articulates the response level uncertainty that is often overlooked by conventional response modes.

Changing the actual format of the data (from traditionally discrete, to interval-valued) is a very substantial shift requiring an equally substantial and sustained research effort across all facets of the process of questionnaire-based research—from capture to analysis. Recent work by present authors has included development of open-source software for efficient digital administration of interval-valued surveys (DECSYS—cf. Ellerby, McCulloch, Young & Wagner, 2019), which is available to download (via https://www.lucidresearch.org/decsys.html) from GitHub. This paper builds upon this—reporting a study designed to provide an initial validation of the underlying approach. In this study participants completed an electronic questionnaire with questions designed to vary systematically along three primary dimensions expected to influence response interval width—inherent variability or range in the appropriate response, adequacy and availability of supporting information, and clarity of question phrasing. We examine whether interval-valued responses reliably reflect induced variability along each of these dimensions—with experimental hypotheses that observed variance in interval widths will be significantly associated with each factor. Results find consistently in the affirmative, providing an initial demonstration of the efficacy of interval-valued responses to systematically capture additional information across a variety of circumstances, relative to discrete response modes.

**The Added Value of Intervals**

Sometimes the correct response to a question is clear and discrete. How many sides does a square have? Here, the answer four is both correct and complete. However, this is not true for all questions. In many cases the appropriate response is uncertain, due to either ambiguity in the question or lack of knowledge of the respondent (cf. Coombs & Coombs, 1976; Converse & Presser, 1986; De Bruin, Manski, Topa & Van Der Klaauw, 2011; Fowler, 1995; Payne, 1951), or because the correct response intrinsically comprises a range of values (cf. Budescu, Broomell & Por, 2009; Harris, Por & Broomell, 2017; Liu & Mendel, 2008; Wagner, Miller & Garibaldi, 2013; Wu, Mendel & Coupland, 2012). These three features are illustrated in the following examples.

First, consider the uncertainty induced by the question 'What number will I roll?' The context of the question is not specified—what is to be rolled? If dice, then how many of them? Of how many sides? And on how many occasions? Viewed alone, the ambiguity in the question makes it impossible to confidently answer. Second, consider the question 'I will roll two fair six-sided dice one time, what sum total will I roll?' Here, the context and the meaning of the question are quite clear. However, the correct answer is inherently unpredictable. Given the stochastic nature of rolling dice, although a best guess can be made, a specific number remains impossible to answer with certainty. The only thing we can say for sure is that the correct answer falls within the (disjunctive) interval [2,12][1]. Third, consider the range of values associated with the question 'I have one standard six-sided die, what numbers are shown on its faces?' The question is worded clearly, and all relevant information is known. However, the answer comprises a (conjunctive) set of distinct values and, to completely express a correct response, the full range of these values must be represented. The minimum complexity response format with the capacity to do so is an interval, with the answer [1,6].

In each of the cases described, any discrete response is insufficient to convey all information necessary for a correct answer. When presented with a discrete response mode, each situation will manifest as response uncertainty—in terms of which single response option represents the most appropriate answer. The act of forcing respondents to condense their answer into a discrete approximation can be viewed as adding noise to the data, as extraneous factors may influence which discrete response option is chosen at the time. Limiting the impact of collapsing a potentially complex response, by enabling interval-valued responses—which intrinsically comprise a range of values (cf. Cloud, Moore & Kearfott, 2009)—should permit the capture of higher fidelity response data. We argue that there are many real-world cases where it would be both more valuable and efficient to offer respondents the opportunity to provide interval-valued responses, rather than currently prevalent point response modes.

One important consideration concerning interval-valued responses is that they may capture different types of information. Specifically, intervals reflecting uncertainty generally comprise a disjunctive set, in the sense that they describe the (lack of) knowledge of the respondent about a specific quantity—as described in the first two cases above—rather than an actual range (cf. Couso & Dubois, 2014; Dubois & Prade, 2012). Confidence intervals are a common example of such disjunctive sets. At the same time, intervals also frequently represent conjunctive sets—i.e., true ranges of values existing in the real world—as described in the third case above. In this paper, we do not focus on defining the nature of the information captured by the interval; we aim rather to demonstrate that capturing intervals can fundamentally afford the systematic and efficient capture of information not available using discrete response modes.

*Practical applications*

One well-documented application for interval-valued judgements is in direct elicitation of uncertainty in respect to individual estimates (i.e., confidence intervals)—whether from experts or otherwise (cf. Alpert & Raiffa, 1982; Cooke, 1991; Hemming, Burgman, Hanea, McBride, &

Wintle, 2018; Kahneman, Slovic & Tversky, 1982; Klayman, Soll, Gonzalez-Vallejo & Barlas, 1999; Morgan, 2014; Soll & Klayman, 2004; Speirs-Bridge et al., 2010). As alluded to above however, it is important to note that the information which can be captured by intervals is not limited to confidence intervals.

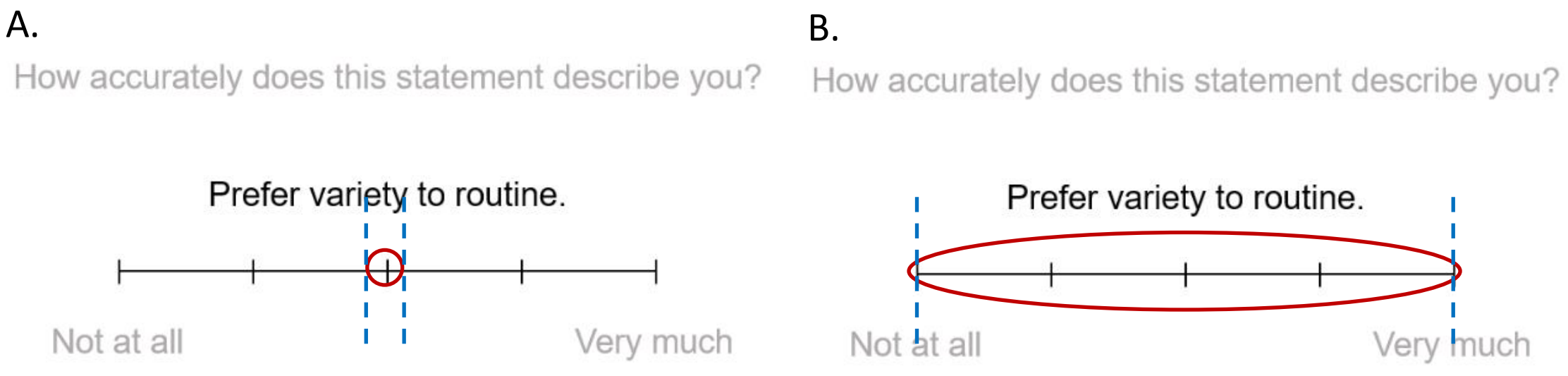

Figure 1: Illustration of how interval-valued responses may be used to differentially indicate neutral and 'Don't Know' responses. 1A shows a neutral response and 1B a 'Don't Know' response, each captured using the 'ellipse technique' where respondents 'circle' the relevant area on the scale. Blue dashed lines indicate interval endpoints extracted from the ellipse extrema, bounded by the ends of the response scale.

In practice, interval-valued responses may be particularly useful in the context of decision research, where participants are routinely asked to make unfamiliar choices and estimate probabilities of outcomes under uncertainty (cf. Cubitt, Navarro-Martinez & Starmer, 2015; Ellerby & Tunney, 2017; 2019, Tversky & Kahneman, 1974). However, aside from eliciting intervals with an explicit degree of confidence, interval size can also be used more generally to identify questions for which participants are more or less uncertain or confident about the most appropriate response (cf. Yaniv & Foster, 1995; 1997). Consider the extreme case of 'Don't Know' responses, which may be expressed by selecting the mid-point of an ordinal scale; interval-valued responses allow for the clear distinction between respondents who are uncertain or ambivalent from those who are genuinely neutral, as shown in Figure 1. Difficulty in doing this is a fundamental limitation of conventional response formats (Coombs & Coombs, 1976;

Klopfer & Madden 1980; Krosnick & Fabrigar, 1997) and has led to the frequent introduction of an additional 'Don't know' response option separate to the scale, increasing complexity for respondents and analysis (discussed further in the next sub-section).

In questionnaire pretesting, interval-valued responses could be used to identify questions that are causing confusion for participants. Here, unusually broad intervals could be interpreted as indicating questions that are ambiguous, imprecise, or otherwise difficult to understand. Complementing existing tools (e.g., QUAID—Graesser, Wiemer-Hastings, Kreuz, 2000), this would allow for easy identification of areas that require improvements in phrasing or presentation, and could also be particularly useful for informing the nature of any observed differences between native and non-native speakers.

In attitudinal surveys, intervals could also be a valuable tool for informing the degree of conviction with which a respondent holds a given attitude or belief, which may be orthogonal to the strength of the attitude or belief itself. For one case—consumer preference research—understanding the degree of conviction or flexibility associated with consumer preferences or attitudes could valuably inform the likelihood of these motivating future behaviour, as well as openness vs resistance to persuasion and potential for establishing consensus (Ellerby, Miles, McCulloch & Wagner, 2020; Petty, Briñol & Tormala, 2002; Rucker, Tormala, Petty & Briñol, 2014).

For another example, consider political science, where facilitating the capture of each respondent's level of commitment regarding their voting intentions could be valuable in informing more accurate models of voting behaviour (cf. Burden, 1997). Specifically, in polling, we could reduce the noise arising from forcing the respondent to collapse their response-model (a complex belief) into a single value at the point of completing the survey. As the same participant may collapse the same belief to a different discrete response when actually voting (e.g., based on immediate external influence, such as a salient news event on the day of polling), it is preferable to capture the actual range of the belief as completely as possible—thus providing not just voter intent at the polling stage, but information regarding the level of voter uncertainty. In fact, a study conducted by Aldrich et al. (1982) positively evaluated the possibility of allowing respondents to select multiple response points along an ordinal response scale. They concluded that their results suggested discrete measures '*force a false precision of response… expressed ambiguity is a real phenomenon. If so, the current practice of forcing respondents to select a single point not only introduces another source of measurement error into already imprecise data, but also loses theoretically important information.'* (p. 411). Nevertheless, discrete response scales predominate to this day.

Finally, intervals will of course be valuable in any case where the most correct and complete answer to a question is itself an interval, such as interpretation of probability phrases (cf. Budescu & Wallsten, 1985; 1995; Karelitz & Budescu, 2004; Harris, Por & Broomell, 2017), or other linguistic terms (Navarro, Wagner, Aickelin, Green & Ashford, 2016; Wu, Mendel & Coupland, 2012). Here, they will inherently improve the fidelity of the response.

**Limitations of Conventional Response Modes**

The putative benefits of any new survey response mode must be considered in relation to existing paradigms. This section focuses on two of the most prevalent, Likert-type and visual analogue scales. Since their development 85 years ago, Likert scales, and in particular the corresponding response format (Likert, 1932), have become established as a ubiquitous method of data collection and basis for analysis over a broad range of research areas. This type of response was designed to capture attitudes; it is therefore commonly used in market-research, to obtain customer feedback, and in psychometric measures such as personality inventories. To be precise, although commonly used to describe the range of allowed responses to a single questionnaire item, the term Likert *scale* refers only to the combined responses relating to multiple items (Carifio & Perla, 2007).

Standard Likert questions require a discrete, ordinal response. These are often selected from a range of 1 to 5, or along a corresponding axis of Strongly Disagree to Strongly Agree. This response format is generally agreed to constitute an ordinal scale, though it is often assumed to represent interval data[2], in the sense that each point is assumed to reflect a value that is equidistant from each adjacent point (Blaikie, 2003). This conjecture has been widely criticised (cf. Bishop & Herron, 2015; Jamieson, 2004; Knapp, 1990; Kuzon, Urbanchek & McCabe 1996), with an important consideration being whether the data in question represents a true scale (i.e., comprising multiple collective responses) or only individual Likert-type questions (Boone & Boone, 2012; Carifio & Perla, 2007; 2008). Nevertheless, counterarguments have quite equivocally proposed that parametric statistics are generally robust for the purposes of analysing Likert-type data, and that their use in these cases does not substantially increase the risk of "*coming to the wrong conclusion*" (Norman, 2010, p. 7), by comparison with non-parametric alternatives.

The standard Likert response format offers relatively low granularity in its most common 5-point form, though this can of course be remediated through increasing the number of response options. Research suggests that doing so has some effect on key data characteristics such as internal consistency, but that this is not transformational (cf. Dawes, 2008; Maydeu-Olivares, Kramp, García-Forero, Gallardo-Pujol, & Coffman, 2009). Alternatively, Visual Analogue Scales (VAS) permit continuously variable responses, by allowing the respondent to select any single point along the represented continuous dimension (cf. Ahearn, 1997; Aitken, 1969; Couper, Tourangeau, Conrad & Singer, 2006; Freyd, 1923). In theory therefore, VAS allow for maximum potential response precision. They have also been shown to approximate an interval-[2] rather than ordinal-scale level of measurement (Reips & Funke, 2008). In practice however, VAS have been found to provide comparable responses to Likert-type scales, while the latter have sometimes been considered preferable in terms of ease of use for the survey respondent (Guyatt, Townsend, Berman & Keller, 1986; Kuhlmann, Dantlgraber & Reips, 2017; van Laerhoven, van der Zaag-Loonen & Derkx, 2004).

However, as alluded to previously, conventional application of these response formats does not consider—or attempt to measure—each individual respondent's uncertainty, or range (e.g., vagueness) inherent in the response. That is, both Likert-type and VAS questions force the respondent to condense their answer to a single-valued datapoint, effectively hiding and resulting in the loss of this associated information. At the data analysis level, statistics concerning between subject (i.e., inter-source) variance can be calculated from discrete responses post-hoc. However, as a method for estimating uncertainty this is relatively inefficient, in that it requires multiple subjects. More importantly, both within-subject (i.e., intra-source) uncertainty and range inherent in each response are fundamentally distinct from this—reflecting heterogeneity within *individual* responses. Capture of this information is the focus of this paper.

It is possible to obtain information on individual response uncertainty using conventional methods. Consider the problem of 'Don't know' responses (cf. Coombs & Coombs, 1976; Klopfer & Madden 1980; Krosnick & Fabrigar, 1997). The inclusion of a separate 'Don't know' option offers one workaround, but with its own fundamental limitations. Uncertainty is not all or nothing, but of varying degrees—when a 'Don't know' option is present, participants are forced to choose between giving a best guess, to communicate their limited information regarding the subject in question, or instead to opt-out, to communicate information only about the presence of some unspecified (though presumably substantial) degree of uncertainty. However, as they cannot express both, in any case where the respondent is neither entirely certain nor entirely ignorant, the response will be incomplete.

Alternatively, dedicated additional questions may be asked about the uncertainty associated with each judgement—which may also be framed in terms of confidence (cf. Peirce & Jastrow, 1884; Cheesman & Merikle, 1984; Dienes, Altmann, Kwan, & Goode, 1995; Sporer et al., 1995; Tunney & Shanks, 2003). However, this approach is inefficient, effectively doubling the number of responses that participants are required to provide. Moreover, although this question is not directly examined in the current study, the lack of cohesion between initial responses and subsequently elicited uncertainty responses may also limit response fidelity—for instance, confidence may differ depending upon response time, and if confidence ratings are provided after initial responses, then these ratings may have continued to evolve in the intervening period (cf. Baranski & Petrusic, 1998; Pleskac & Busemeyer, 2010). Finally, although they often correspond, we note that 'confidence' is a complex notion inequivalent to (un)certainty—and reported confidence may have distinct determinants from uncertainty reported through interval width (Teigen & Jørgensen, 2005). For example, a respondent may be perfectly confident of an inherently interval-valued response (e.g., the conjunctive set of numbers shown on a standard die, as mentioned earlier).

**Enabling Efficient Interval Capture Using Ellipses**

An emerging ellipse-based response mode has been developed to allow efficient, coherent, and intuitive capture and quantification of uncertainties and ranges associated with individual survey responses. This method is designed to leverage the growing prevalence and public familiarity with modern information technologies—such as accurate and portable touchscreen devices—and builds upon widespread familiarity with 'circling' areas of interest to minimise, and potentially eliminate, any added effort investment required to obtain this richer response information. It is designed as an efficient compromise between traditional and currently ubiquitous methods, such as discrete ordinal scales, and more complex or esoteric approaches, such as qualitative interviews, various methods of eliciting (specifically) probability distributions (cf. De Bruin et al., 2011; Morris, Oakley & Crowe, 2014; Speirs-Bridge et al., 2010), including SHELF (Gosling, 2018; O'Hagan, 2019), and the 'Fuzzy Graphic Rating Scale' (FRS), (cf. Hesketh, Pryor & Hesketh, 1988; Hesketh, McLachlan & Gardner, 1992; Lubiano, de Sáa, Montenegro, Sinova & Gil, 2016; Quirós, Alonso & Pancho, 2016).

By comparison with currently predominant Likert-type responses, the ellipse response mode replaces ordinal options with a continuous scale, akin to that of the VAS. However, participants provide an interval-valued, rather than point, response. While a variety of approaches could be considered to enable provision of interval-valued responses, in this case respondents do so by drawing an ellipse on the scale, delineating the interval that they believe best represents the answer. Crucially, participants are encouraged to use the width of this ellipse (i.e., the size of the interval) to indicate the degree of range (e.g., uncertainty, variability, vagueness, or ambiguity) in their response. This is illustrated in Figure 2, by comparison with more traditional Likert-type and VAS response formats.

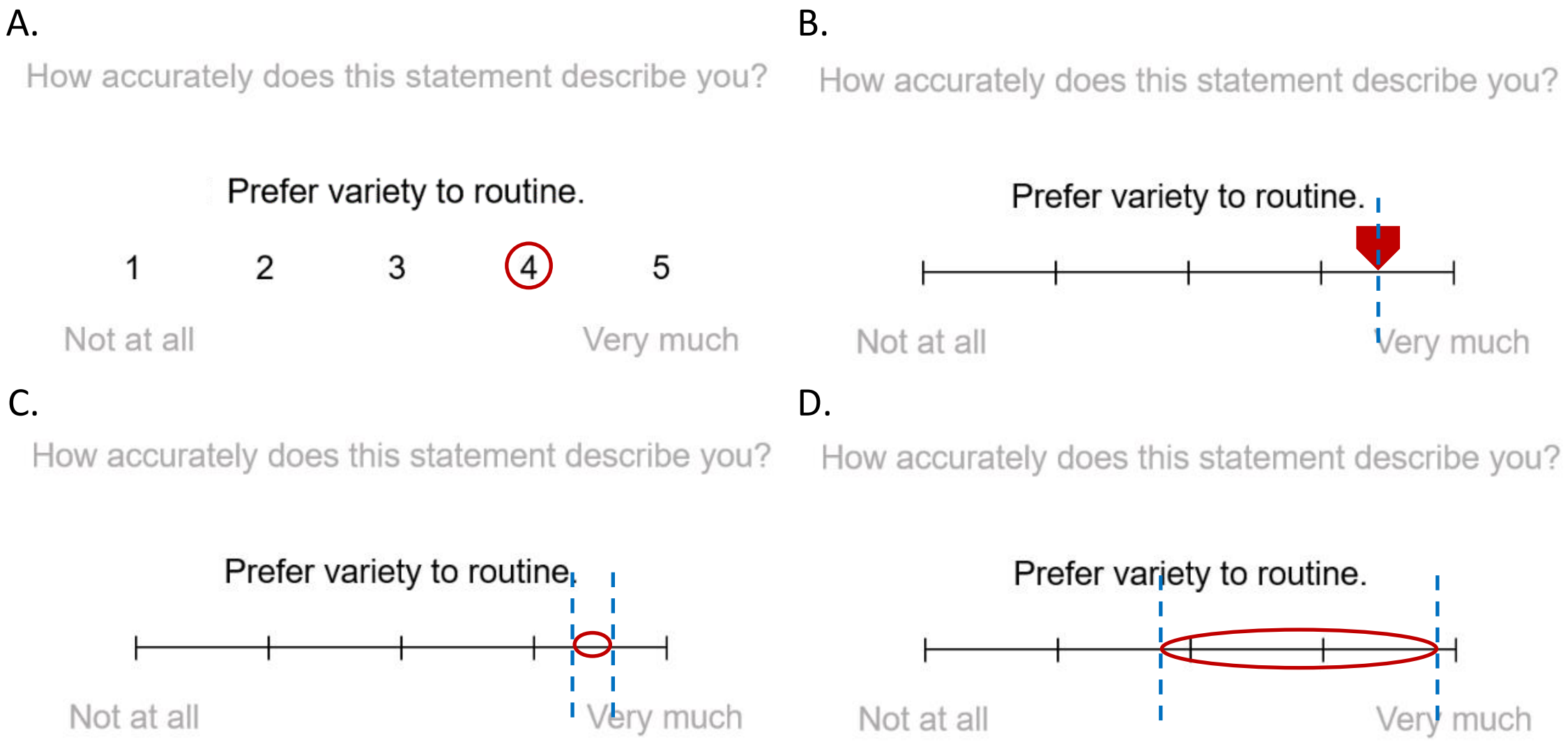

Figure 2: Example questionnaire responses. 2A shows a Likert-type ordinal response. 2B shows VAS-type response. 2C shows example interval-valued response with low uncertainty. 2D shows an example interval-valued response with high uncertainty. Divisions (sub markers) on continuous scales are illustrative here only and questions of appropriate scale design apply as for traditional scales.

Therefore, this interval-valued response mode extends beyond the benefits of existing continuous scales, such as VAS. It has the potential not only to capture responses that are precise, but also makes possible the capture of fundamentally distinct information, such as individual response uncertainty or vagueness. This information is usually lost when using point-valued response formats. Moreover, ellipses achieve this in a single, integrated, and cohesive response. These attributes should make the approach intuitive to use and quickly and easily administrable, facilitating broader adoption while reducing training requirements necessary to obtain high fidelity responses.

The ellipse approach is designed primarily to streamline the process of interval-valued data collection so far as to provide an alternative to point-valued response modes (e.g., Likert-type or VAS), where it offers a substantial informational advantage at a minimally increased, or potentially even reduced, workload—e.g., by counteracting choice paralysis when selecting between multiple potentially appropriate discrete alternatives or requiring fewer questions to be asked. On the flipside, it is not designed to directly compete with other approaches that aim to maximise information obtained concerning uncertain responses, at the cost of substantially increased time and effort in administration—considering training requirements as well as the data collection process itself.

Previous research on the elicitation of confidence intervals (an example of disjunctive sets) has consistently indicated that interval-valued estimates tend to be overconfident (Alpert & Raiffa, 1982; Juslin, Wennerholm & Olsson, 1999; Klayman, 1999; Yaniv & Foster, 1995; 1997). Further evidence suggests that singular interval estimates, such as those elicited by the ellipse response mode, are prone to greater overconfidence than those split into multiple steps (Soll & Klayman, 2004; Speirs-Bridge et al., 2010). At the same time, the coherent, singular nature of ellipse responses is a key asset in avoiding the added complexity and effort induced by soliciting multiple responses for each survey item.

While this goes beyond the scope of the present paper, there is evidence to suggest that for uncertainty capture (i.e., the disjunctive case), ellipses may be better placed to capture respondents' subjectively 'reasonable' bounds—with the potential for follow-up questions to determine subjective confidence in the accuracy of the interval provided (Teigen & Jørgensen, 2005; Winman, Hansson & Juslin, 2004), and potentially calibrate or convert these into standardised credible intervals or probability distributions (cf. Hemming et al., 2018; Speirs-Bridge et al., 2010). However, as discussed earlier, confidence intervals and, more generally, disjunctive sets are only one example of the information which can be captured by intervals. In practice, the most appropriate method to use will depend upon both the questions to be answered and the time and effort available to invest. Here we focus on the ellipse response mode as an alternative to day-to-day surveys that use discrete responses.

In different disciplines, related research has already been conducted into practical foundations for using interval-valued responses. Much of this has focused on the exploration and development of appropriate mathematical methods to effectively extract, process and interpret the resulting data (cf. Anderson et al., 2014; Liu & Mendel, 2008; McCulloch, Ellerby & Wagner, 2019; 2020; Havens, Wagner & Anderson, 2017; Miller, Wagner & Garibaldi, 2012; Wagner, Miller, Garibaldi, 2013; Wagner, Miller, Garibaldi, Anderson & Havens, 2015; Wu, Mendel & Coupland, 2012)—leveraging existing theory from the fields of Interval Arithmetic (cf. Cloud, Moore & Kearfott, 2009; Moore, 1966; Nguyen, Kreinovich, Wu & Xiang, 2012) and Fuzzy Set Theory (Zadeh, 1965). The present authors have also directed substantial recent work into developing a practical software tool (DECSYS—Ellerby et al., 2019; 2020), which now makes efficient digital collection of interval-valued survey data scalable and widely accessible.

Nonetheless, despite the promise of and growing interest in this response mode, it remains to be conclusively demonstrated whether interval-valued responses provided through ellipses reliably capture set-valued information (either disjunctive—e.g., arising from uncertainty, or conjunctive—e.g., arising from vagueness); or whether, due to either participant satisficing or careless responding (cf. Krosnick & Alwin, 1987; Meade & Craig, 2012), they might instead be arbitrary or dependent upon other extraneous factors. The present study was designed to address this gap in the evidence base and determine whether this 'quick and easy' approach can indeed provide robust and valuable additional information associated with individual responses—establishing a foundation for further research and development into methods for efficient interval capture.

**Experimental Summary**

The aim of the experiment was to assess the efficacy of interval-valued survey responses to capture meaningful additional information, across a variety of circumstances and relating to three potentially distinct sources, which may each be present when responding to conventional survey questions. Over a three-section digital questionnaire, we induced varying degrees of each: inherent variability in the appropriate response, adequacy and availability of supporting information, and clarity of question phrasing—to determine whether the variance in size of elicited intervals is systematic, with reference to established ground-truth in each of these cases.

In section one, participants were asked to simply identify and reproduce data from a chart, which itself comprised a range of specific values (i.e., a conjunctive set). The objective here was simply to establish whether respondents, given minimal instruction, are able to understand the basic concept of an interval-valued response, and to use the ellipse format to communicate ranges that they are aware of—i.e., which were in this case explicitly provided.

In section two, participants made their own judgements concerning stimuli with experimentally controlled degrees of both uncertainty and variability. The aim here was to examine how respondents apply the interval-valued response format to represent their own experienced degrees of stimulus-related uncertainty and variability (i.e., both disjunctive and conjunctive aspects), as well as the interplay between these two factors.

In section three, participants responded to a broader variety of items, which were designed to differ systematically in levels of ambiguity, specificity, and comprehensibility. In this section we explore how respondents use interval-valued responses to communicate their own subjectively perceived ambiguities or degrees of uncertainty, when induced by lack of clarity in, or comprehension of, the question or associated stimuli.

Eliciting interval-valued responses from participants in each of these cases, following only brief instruction regarding how to use this response format at the outset of the study, should provide insight into the potential for practical utility and real-world adoption of this response mode across a range of contexts and situations.

## Method

### Participants

A total of 40 participants completed this experiment, recruited in an opportunity sample. As the study was advertised primarily across University of Nottingham UK Campuses, via physical posters and email list invitations, members of the university community were disproportionately represented. Insofar as is feasible within this population, the sample was varied—with participants originating from a variety of schools, UK campuses and job roles (a substantial proportion of staff participants were non-academic). Each participant was paid a fixed inconvenience allowance of £5 (GBP), to compensate for their time spent taking part. Nineteen of these self-identified as male and nineteen female, two declined to report their gender. Ages ranged from 18 to 50 (M=25.65, SD=7.75). Twenty-four reported as native English speakers, fifteen as non-native speakers, while one declined to report this.

A range of different statistical analyses were conducted, statistical power will differ substantially between these. For the simplest cases, i.e., one-sample and paired samples parametric *t*-tests, this final sample size was determined to offer a statistical power of >.99, .87 and .23, to detect a large, medium or small effect respectively (two-tailed, $\alpha$ = .05, d=.8, .5, & .2—cf. Cohen, 1988). For ANOVAs, power depends heavily upon specifics—including number of measurements, as well as factors that are not known a priori, such as correlations among repeated measures and sphericity of the data to be obtained. Nonetheless, calculations using default values for these parameters indicated, across all planned ANOVAs, a minimum power of >.99 to detect a large effect, .94 to detect a medium effect and .26 to detect a small effect ($\alpha$ = .05, corr. among rep. measures = .5, sphericity assumed, *F*=.4, .25, .1 respectively—cf. Cohen, 1988). When applying worst case scenarios for sphericity, power across all analyses remained at minimum .95, .62 and .15, for *F*=.4, .25 and .1. All power calculations were made using G*Power (Faul, Erdfelder, Lang & Buchner, 2007).

## Questionnaire Items

All questionnaire items were developed for the purposes of this study. These were classified into three distinct sections. The purpose and content of all three sections was briefly explained to each participant before the beginning of the study, and this was reiterated for each individual section before it was begun.

In section one, participants observed a chart showing claimed natural life expectancies for several different household pets, if given proper care and attention (see Fig. 3). This was created using data obtained from *injaf.org*, under a Creative Commons agreement for non-commercial use, and with written permission from the copyright holder. The data provided were already in the form of intervals (e.g., ranging between 15 and 20 years for a domestic cat)—the task of respondents was therefore simply to reproduce these data as best they could, over a series of eight specific animals, using the interval-valued scale.

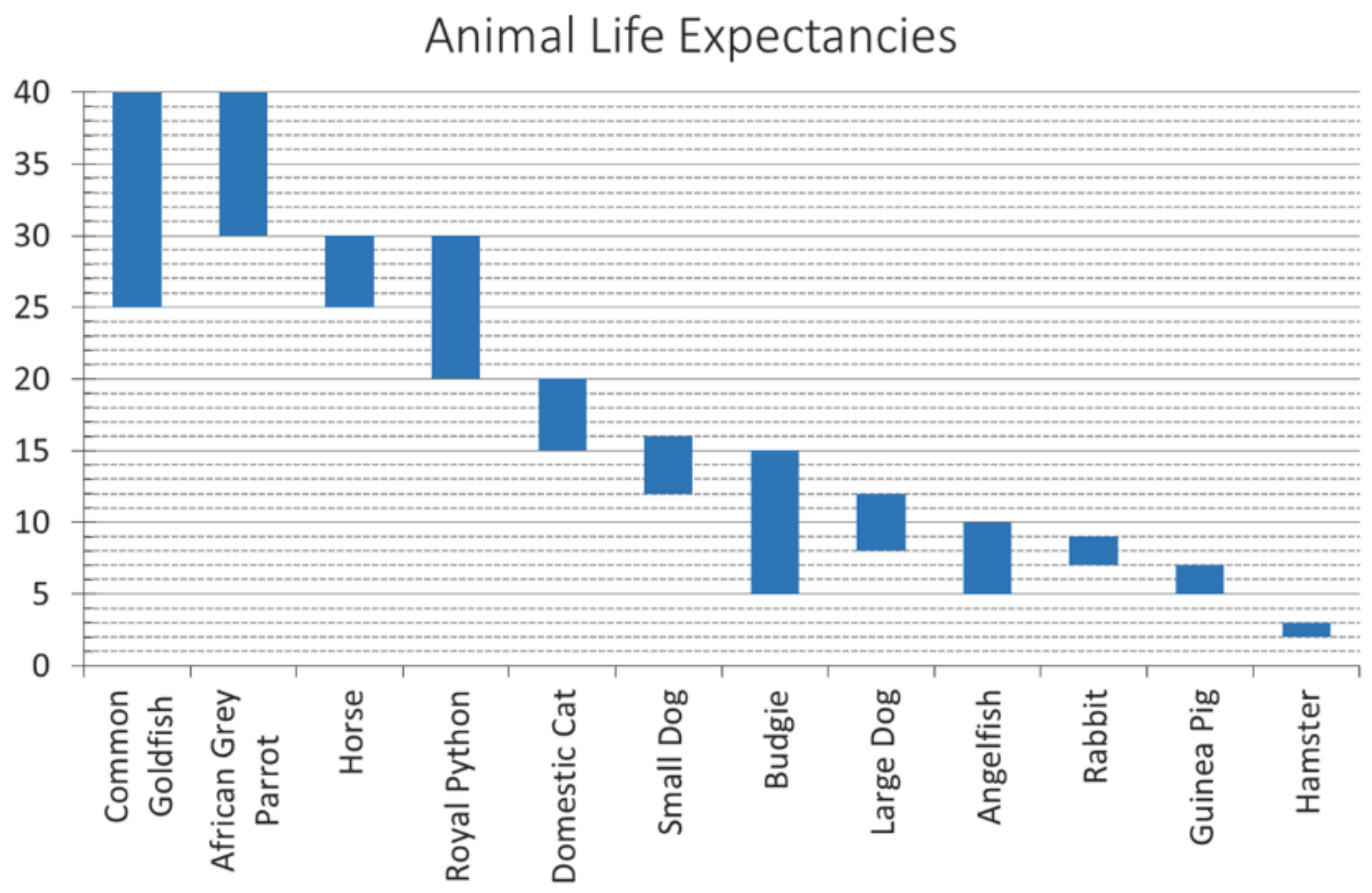

Figure 3: Showing animal life expectancy reference chart, from the first section of the survey.

In section two, participants were presented with a series of six sets of 35 marbles—each comprising some proportion of blue versus yellow marbles. Each set was organised into five rows, of seven marbles each. Respondents were tasked with providing their best estimate for how many blue marbles were present overall in *each row* of the given set—with a response scale ranging between zero and seven. Initially, for each set, six out of seven marbles in each row were hidden, entailing a substantial degree of uncertainty around this estimate. In the following questions, three more, and then all marbles in each row of the set were revealed. This process was repeated for each of the six sets, for a total of 18 questions. Importantly, for some sets the number of marbles in each row was identical, but for others there was a discrepancy between rows. This between row variability entailed a degree of range in the correct interval-valued response. Example stimuli are shown in Figure 4.

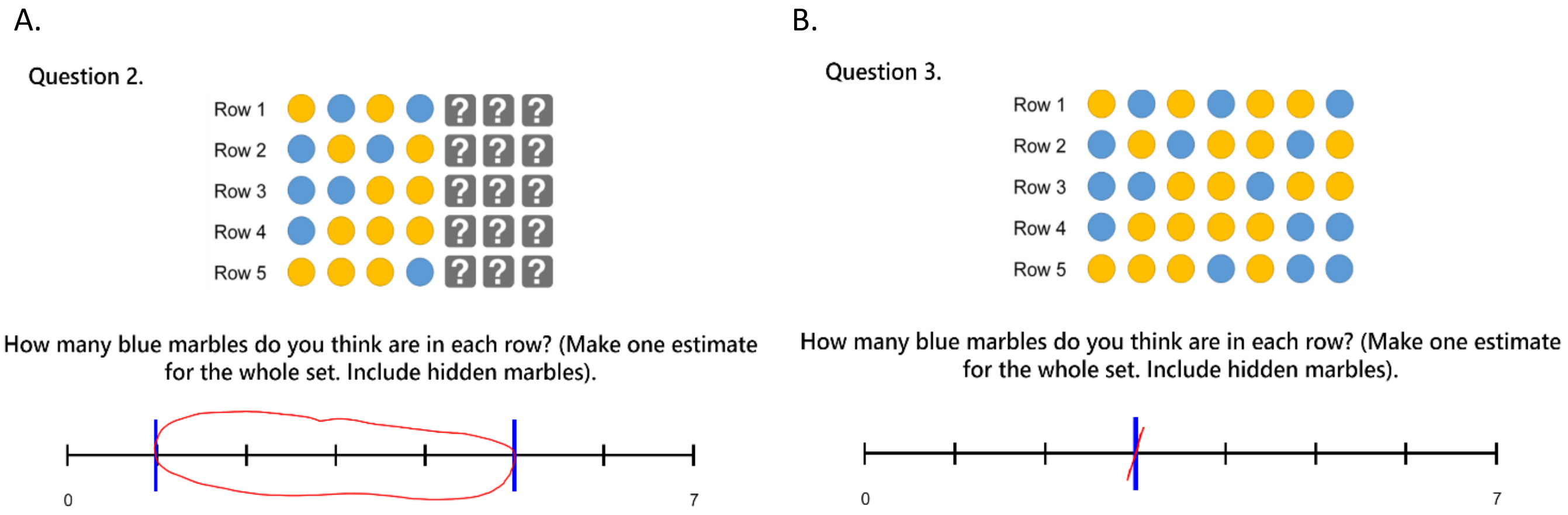

Figure 4: Showing example marble stimuli, from the second section of the survey. In 4A, three marbles from each row remain hidden. In 4B, all 35 marbles are visible. Example (expected) responses are shown below each stimulus.

In the third and final section, respondents answered a collection of more varied items. These questions were designed to hold varying degrees of vagueness, ambiguity or comprehensibility in their wording or meaning. Crucially, although questions were deliberately worded to induce qualitatively more or less uncertainty, ambiguity or vagueness in the response, the precise quantity of these depended upon the subjective interpretation of the respondent. First, questions were asked concerning the temperature in England, either during a specific month (*December, July*), or without qualification; the aim being to establish whether interval widths could be used to distinguish between more and less specifically phrased questions. Second, questions were asked concerning a scenario in which a percentage value either *increased by 50%* or *doubled*—with the former being possible to interpret as either a 50 percent or 50 percentage point increase on the original amount. The aim here was to determine whether interval widths could discriminate questions with clear vs potentially ambiguous interpretations. Third, respondents were asked to report how well they were described by a series of twelve personality-related words—four of these were commonly used (*talkative, aggressive, lazy, quiet*), four were low frequency (*garrulous, bellicose, indolent, taciturn*), and four were not real words, but created solely for use in the study (*brendacious, apoccular, lombardistic, revenotile*). The objective here was to establish whether intervals could identify questions within which a word was poorly, or not, understood. Fourth, respondents were asked a series of double-barrelled questions, followed by their single-barrelled counterparts (e.g., How well does this statement describe you?—I like *reading books and watching television*, vs I like *reading books*, vs I like *watching television*. Also: w*atching and playing sports, drinking tea and coffee, cooking and eating*). The objective here was to further establish whether intervals could discriminate more and less specifically phrased questions.

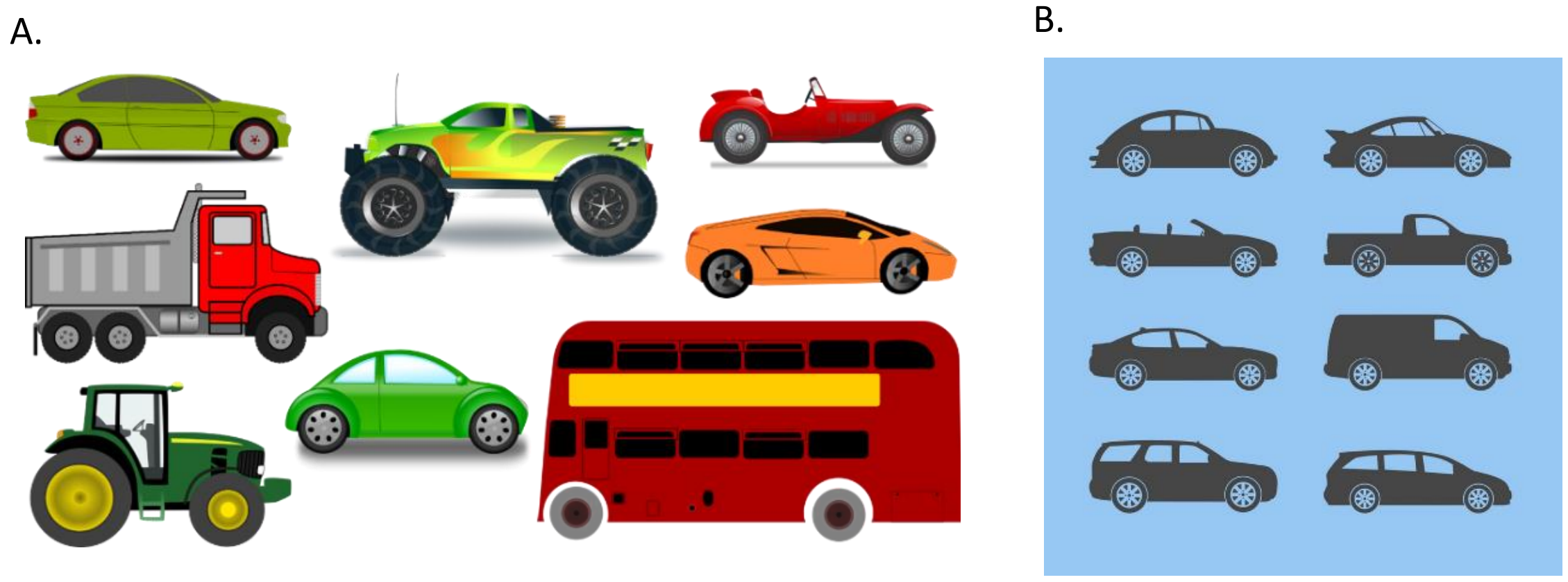

Figure 5: Showing example vehicle stimuli, from the third section of the survey. Both stimuli contain a total of eight vehicles, but a potentially disputable number of *cars*.

The final two question types in this section moved the focus from the question wording alone, to the combination of question and associated stimulus. Fifth, respondents were asked to judge how many *cars*, and then how many *vehicles*, were present within each of two images (see Fig. 5)—each of these contained eight vehicles (the more inclusive term) in total, of which a potentially disputable number could be considered cars. Sixth, and finally, respondents were asked to judge the number of blue marbles contained in three rows, of eight marbles each (see Fig. 6). For the first of these, the correct response was clear, with marbles each being either blue (4), green (2), or yellow (2)[3,4]. For the next, a gradient was introduced from blue to green to make the number of blue marbles blurred, or vague. For the final set, marbles each contained only blue or yellow[3], but five marbles were patterned, containing varying proportions of each colour. As the question did not clearly specify whether the term blue was intended to refer only to completely blue, or also to mainly, or even partially blue marbles, the appropriate response could be considered ambiguous. The aim in each of these cases was to further examine whether the width of interval-valued responses would enable discrimination between question-stimulus combinations that were clear, and those that may be interpreted as vague, unclear or ambiguous.

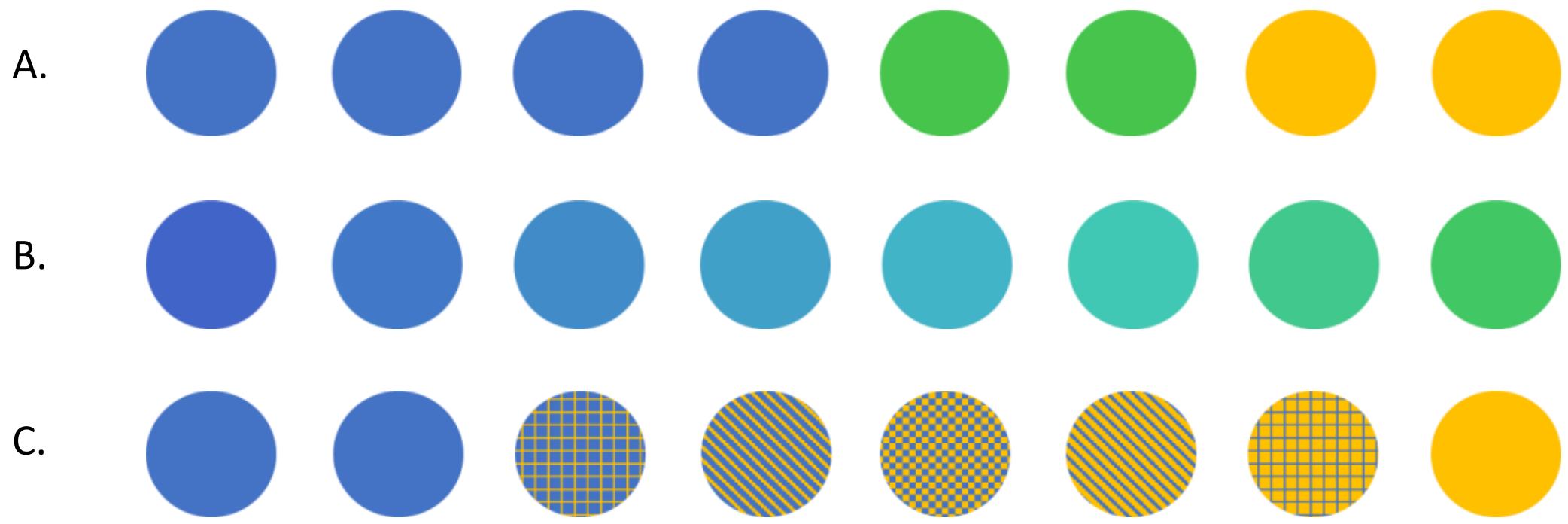

Figure 6: Showing example marble stimuli, from the third section of the survey. In 6A, the number of blue marbles is clear. In 6B, a gradient blurs the transition from blue to green. In 6C there is a lack of clarity in whether *blue* refers to entirely, mainly, or partially blue marbles.

In the subjective feedback section of the questionnaire, participants were asked four questions. These were whether they found the response format '*easy to use*', '*unnecessarily complex*', and whether it allowed them to '*effectively communicate* [their] *desired response*', as well as whether '*Overall,* [they] *liked the response format*'. These were administered using a conventional 5-point ordinal scale, ranging from *'Strongly Disagree'* to *'Strongly Agree'*.

**Experimental Design**

The study used a repeated measures design, in which each participant completed the entire 63-item questionnaire. All responses were made through the ellipse-based interval-valued response mode, using a touchscreen Microsoft Surface Pro computer, together with a stylus. All surveys were administered through a local area network, using the DECSYS software's *Workshop Mode* (Ellerby et al., 2019)[5]. In this paper, response intervals were deconstructed into two separate dependent variables, namely their position (captured by the interval mean) and width (the distance between interval endpoints)[6,7]. Participants also completed a short feedback questionnaire—adapted from the Systems Usability Scale (Brooke, 1996)—for which the dependent variable was degree of agreement, represented along a 5-point ordinal response scale.

Question order was not randomised for each respondent, due to design factors combined with limitations to the randomisation capabilities of the early version of DECSYS used for the study. In Section 2 for instance, it was necessary to present the three questions relating to each set of marbles together and in sequence, with six, then three, then no hidden marbles. Question order must therefore be considered as a potential factor when interpreting results; for example, participants had more experience providing interval-valued responses by the time they were completing the latter stages of the survey.

**Procedure**

Prior to the study, ethical approval was obtained from the University of Nottingham, School of Computer Science—where the lead authors are based. Before beginning the experiment, each participant was presented with a project information sheet, and informed consent was obtained. All participants were free to withdraw from the study at any time and without giving a reason. Following this, participants were individually seated before a Microsoft Surface Pro touchscreen computer and began the survey process. Initially, participants read a general instructions page, in which the appropriate use of the interval-valued questionnaire response format was briefly explained to them (see Appendix A). This instructed participants to use an ellipse to mark each answer, and that a wider ellipse should be used to indicate greater uncertainty, range, or vagueness in the desired response. Illustrated examples were provided here of both a more and a less certain response (see also Figure 2C, 2D). After this, participants read through three slides of task instructions, which provided basic information regarding each section of the survey (see Appendix B). Participants were then asked to indicate whether they had read and fully understood the task information, and to raise any outstanding questions with the experimenter. Once they had done so, they proceeded to complete all three sections of the survey—questions are detailed in the *Questionnaire Items* section. Total task duration was generally 30-40 minutes. Following completion of the main survey, participants were asked to complete a few more short questions, this time on paper, designed to elicit their subjective feedback about the interval-valued response format. Once these were completed, participants were informed that they had finished the study and provided with their inconvenience allowance.

# Results

A series of statistical analyses were conducted, relating to response-data obtained within the different sections of the survey. These are detailed in the following sub-sections.

It is important to bear in mind that for the purposes of the analyses performed here we extract only certain characteristics from the intervals (i.e., *interval position*—as the mean of the interval endpoints, and *interval width*—as the difference between the left and right endpoints)[6,7].

## Section 1: Reproducing Presented Intervals

The purpose of this section was to assess the capability of respondents to reproduce intervals explicitly provided to them, using the ellipse response mode. Visualisations of the interval-valued responses provided in relation to two of the items in this section are shown in Figure 7; one shows the raw intervals provided by each respondent, and the second represents an aggregation of these responses, based on the Interval Agreement Approach (IAA—cf. Wagner et al., 2015). The IAA represents the degree of agreement across the group—i.e., overlap between all intervals—as a 2-dimensional distribution. This may be interpreted as a fuzzy set (cf. Zadeh, 1965), with the agreement between intervals at each *x*-value determining the corresponding degree of membership.

For comparative analysis, we determined both Pearson correlation coefficients (*r*-values) and mean square error values (M.S.E), which were based on a standardised response scale over [0,100] (to permit comparison Section 2 results). This was done for both the midpoints and widths of the intervals drawn by each of the 40 respondents, over all eight questions, by comparison with those originally provided on the animal lifespan chart (Fig. 3). The resulting groups of 40 *r*-values (representing each respondent) were then compared against zero using one-sample *t*-tests (two-tailed), to examine the group-level relationship between stimulus and response intervals. As these data were consistently found to significantly deviate from normality, we used bootstrapping to ensure robustness (10,000 samples). We also report M.S.E values with bootstrapped 95% confidence intervals.

These tests revealed a significant and strong positive correlation between original and drawn interval midpoints, $p$=.003 (M = .97, 95% CIs .93 to 1.00). Also, between original and drawn interval widths, $p$=.002 (M = .89, 95% CIs .78 to .97). M.S.E. values—with scale range standardised to [0,100]—were as follows: for interval midpoints M = 55.29 (95% CIs 6.70 to 122.27), for interval widths M = 33.72 (95% CIs 8.99 to 65.65). These results indicate that the interval-valued responses provided in this section closely approximated the original (ground truth) intervals upon which they were based—in terms of both position and width.

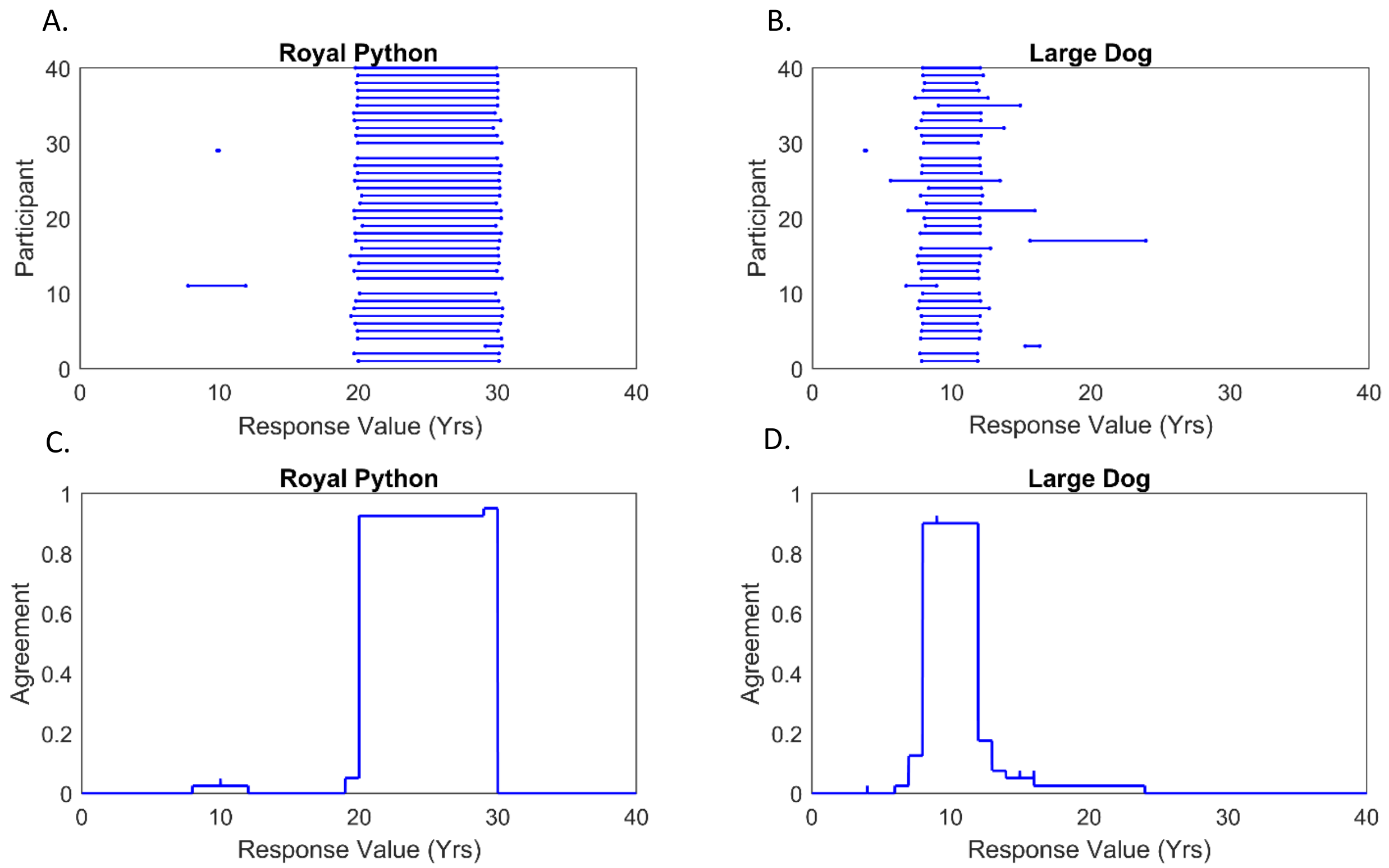

Figure 7: Responses to questions regarding the natural lifespan of a Royal Python and Large Dog (stated as 20-30 and 8-12yrs—see Fig. 3). Panels A & B—showing all interval-valued responses. Panels C & D—IAA plots showing aggregated interval-valued responses (rounded to nearest response integer).

**Section 2: Reporting Disjunctive and Conjunctive Intervals**

The purpose of this section was to examine whether participants were able to use the ellipse response mode to generate intervals representing observed degrees of both uncertainty and variability—and then to assess just how the intervals provided reflect these two factors. Two independent forms of analysis were applied to achieve these objectives. First, the objectively definable maximum and minimum bounds on the possible number of blue marbles in each row were determined for each question, as a benchmark (ground truth) for comparison. Then, as in the previous section, midpoints and widths of these intervals were compared against the corresponding responses provided by each of the 40 participants, to obtain both *r* and M.S.E values (based on a standardised response scale from [0,100]) for each respondent across all questions. As in Section 1, these data were found to significantly deviate from normality, so *r*-values were compared against zero using bootstrapped one-sample *t*-tests (two-tailed, 10,000 samples), and M.S.E values are reported with bootstrapped 95% confidence intervals.

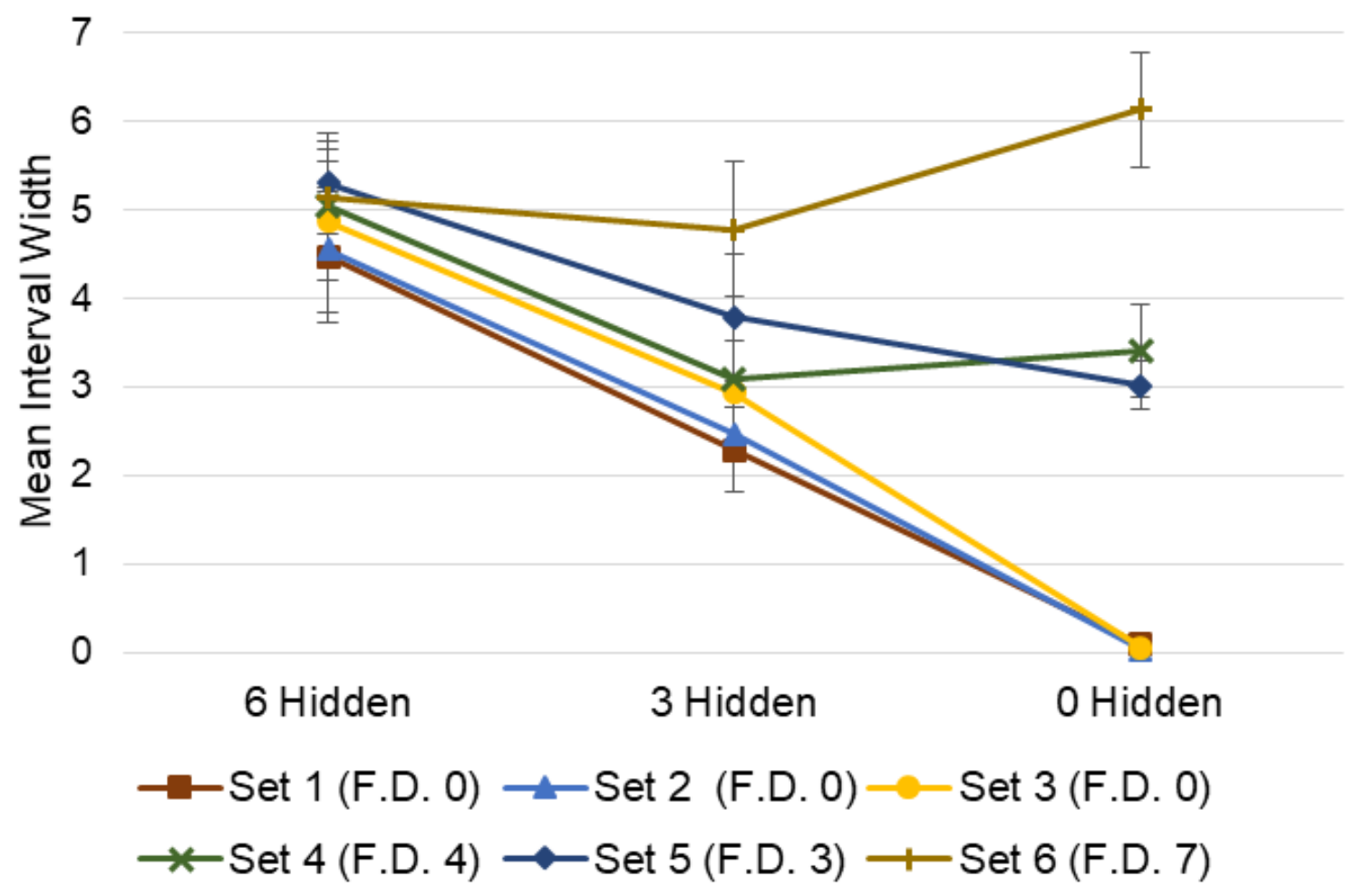

Figure 8: Showing mean interval widths for each Section 2 stimulus, to illustrate the combined effects of withheld information (hidden marbles), and stimulus variability (discrepancy between rows), on the widths of response intervals. 'F.D' (Final Discrepancy) represents the maximum discrepancy in the number of blue marbles between any two rows of marbles in the 'final' set—i.e., once all marbles are visible. Error bars show 95% CIs.

These tests revealed a significant and strong positive correlation between benchmark and drawn interval midpoints, $p<.001$ (M = .90, 95% CIs .861 to .937), as well as between benchmark and drawn interval widths, $p<.001$ (M = .75, 95% CIs .683 to .820). In this case, M.S.E. values were: M = 65.72 (95% CIs 37.60 to 99.52) for interval midpoints, and M = 954.01 (95% CIs 663.76 to 1267.01) for interval widths. These results indicate that the interval-valued responses provided in this section were again strongly associated with expected values—in this instance representing minimum and maximum bounds on the possible range of blue marbles, across all rows in any given set. However, in contrast with the previous section, where the task was simply to reproduce intervals, deviation from benchmark interval width values was substantially greater.

*Mixed effects modelling*

This second stage of analysis applied linear mixed-effects modelling to examine how midpoints and widths of respondent provided intervals were influenced by three salient factors that were visible to each respondent—two being putative sources of interval width, representing disjunctive and conjunctive aspects, respectively. These factors were the proportion of visible marbles that were blue, the proportion of marbles that were hidden (i.e., epistemic uncertainty—disjunctive), and the discrepancy in the visible number of blue marbles between rows (i.e., ontic range—conjunctive)—this was proportional to the maximum possible discrepancy, of seven. This analysis was conducted separately for the dependent variables of midpoint and width—each also represented as a proportion of the size of the entire response scale.

These three factors were entered into each model, alongside three two-way interaction terms. Two random intercepts were also incorporated, one for participants and another for questionnaire items. These allow the models to account for differing baseline positions and widths between subjects and between items. Models were estimated using the Restricted

Maximum Likelihood method, by means of the *fitlme* Matlab function. The model formula to explain interval midpoint ($\gamma_{i,j}^{m}$) and interval width ($\gamma_{i,j}^{w}$), captured by $\gamma_{i,j}^{z}$ was therefore as follows:

$$\gamma_{i,j}^{z} = b_0^z + b_1^z x_{i,j}^{B} + b_2^z x_{i,j}^{H} + b_3^z x_{i,j}^{D} + b_4^z x_{i,j}^{B} * x_{i,j}^{H} + b_5^z x_{i,j}^{B} * x_{i,j}^{D} + b_6^z x_{i,j}^{D} * x_{i,j}^{H} + \mu_i + \mu_j + \varepsilon_{i,j}$$

Where $z$ represents the outcome variable, which may be either midpoint (*m*) or width (*w*), for participant $i$ on item $j$. $b_0$ denotes the fixed intercept, while $\mu_i$ and $\mu_j$ denote respective random intercepts for participant and questionnaire item. The remaining $b$ terms denote the coefficients of the six respective fixed effects: amount of visible blue marbles per row $x^B$, amount of hidden marbles per row $x^H$, the size of the discrepancy in blue marbles between rows $x^D$, plus two-way interaction terms. $x_{i,j}$ represents the observation for each participant and item, and $\varepsilon_{i,j}$ the error term. Results are shown in Tables 1 and 2.

Table 1: Showing effects of salient factors on position (midpoint) of response intervals.

| *Fixed Effects Estimates: DV Midpoints* | *b* | *SE* | *95% C.I.* | *t* | *p* |
|---|---|---|---|---|---|
| Fixed: | | | | | |
| Intercept | .006 | .014 | -.022, .034 | .437 | .662 |
| Blue Visible – ($x_{i,j}^{B}$) | .979 | .030 | .920, 1.037 | 32.785 | <.001 |
| Hidden – ($x_{i,j}^{H}$) | .494 | .024 | .446, .542 | 20.238 | <.001 |
| Row Discrepancy – ($x_{i,j}^{D}$) | .389 | .036 | .319, .459 | 10.911 | <.001 |
| Blue V. * Hidden – ($x_{i,j}^{B} * x_{i,j}^{H}$) | -.985 | .047 | -1.078, -.893 | -20.946 | <.001 |
| Blue V. * Row D. – ($x_{i,j}^{B} * x_{i,j}^{D}$) | -.472 | .091 | -.652, -.293 | -5.163 | <.001 |
| Row D. * Hidden – ($x_{i,j}^{D} * x_{i,j}^{H}$) | -.145 | .071 | -.285, -.005 | -2.034 | .042 |
| *Random Effects Estimates* | $\mu$ | | | | |
| Participant Intercept – ($\mu_i$) | .018 | | .011, .031 | | |
| Question Intercept – ($\mu_j$) | .008 | | .002, .041 | | |
| Residual – $\varepsilon_{i,j}$ | .089 | | .084, .093 | | |

Number of Observations = 720, AIC = -1365.9, BIC = -1320.2

Regarding the midpoint of participant estimates, results indicate that this increased in direct proportion with the amount of visible blue marbles, and at a 1:2 ratio with the number of hidden marbles—consistent with participants assuming a 50% likelihood of each hidden marble being blue. The two-way interaction term between these two variables closely compensates for overestimation in cases where there is both a high proportion of visible blue marbles and a high proportion of hidden marbles (i.e., this effect regulates the effect of proportion of visible marbles that were blue according to the proportion of total marbles that were visible, such that the midpoint increases in line with proportion of total marbles that were blue). Interestingly, a greater discrepancy between rows also leads to higher average estimates, although this effect is ameliorated in cases where there is a high proportion of either hidden or visible blue marbles.

Table 2: Showing effects of salient factors on size (width) of response intervals.

| *Fixed Effects Estimates: DV Widths* | *b* | *SE* | *95% C.I.* | *t* | *p* |
|---|---|---|---|---|---|
| *Fixed:* | | | | | |
| Intercept | .004 | .055 | -.105, .112 | .069 | .945 |
| Blue Visible – $(x^B_{i,j})$ | .012 | .104 | -.191, .216 | .119 | .906 |
| Hidden – $(x^H_{i,j})$ | .783 | .085 | .616, .950 | 9.222 | <.001 |
| Row Discrepancy – $(x^D_{i,j})$ | .994 | .124 | .751, 1.237 | 8.031 | <.001 |
| Blue V. * Hidden – $(x^B_{i,j} * x^H_{i,j})$ | .005 | .163 | -.315, .327 | .034 | .973 |
| Blue V. * Row D. – $(x^B_{i,j} * x^D_{i,j})$ | -.335 | .318 | -.959, .289 | -1.053 | .293 |
| Row D. * Hidden – $(x^D_{i,j} * x^H_{i,j})$ | -.888 | .247 | -1.374, -.402 | -3.590 | <.001 |
| *Random Effects Estimates* | $\mu$ | | | | |
| Participant Intercept – $(\mu_i)$ | .165 | | .130, .209 | | |
| Question Intercept – $(\mu_j)$ | .047 | | .026, .086 | | |
| *Residual* – $\varepsilon_{i,j}$ | .199 | | .189, .211 | | |

Number of Observations = 720, AIC = -120.5, BIC = -74.8

Regarding the widths of participant estimates, results indicate that these increased in direct proportion to the observed discrepancy between rows, but at only an approximately 4:5 ratio with the number of hidden marbles. Note that the upper 95% CI of the latter effect is below one—indicating that this tendency to underestimate uncertainty concerning hidden marbles was statistically significant. The two-way interaction term between these two variables indicates that the combined effect of hidden marbles and visible discrepancy is also less than the sum of its parts. This is consistent with two potential explanations: First, that respondents find it difficult to integrate conjunctive and disjunctive sources of interval width—i.e., the presence of between row variability detracts from accurate uncertainty reporting, exacerbating the general tendency to underestimate this. Second, that when more marbles are hidden from view it is reasonable to give less weight to observed discrepancies, as these may yet 'balance-out' when the hidden marbles are revealed—but as more marbles become visible it becomes increasingly likely that any visible discrepancy will be reflected in the final set. All effects relating to the proportion of blue visible marbles were found not to have any significant effect on response widths.

**Section 3: Generating Subjective Intervals**

The purpose of this section was to examine how respondents used intervals to describe and communicate their own subjectively perceived uncertainty about, or range inherent in, the appropriate response, arising from the phrasing or subject of a given question. For this section, a variety of statistical analyses were applied, as appropriate, to examine differences in responses between items within each of the different sub-sections. Note that ANOVA model residuals were consistently found to violate the normality assumption. Therefore, we also performed non-parametric robustness checks in each case, based upon Monte-Carlo resampling (10,000 samples). Results were found to match the parametric test outcome in every instance, and are detailed in Appendix C.

First, we assessed questions deliberately worded in a more or less specific manner—relating to subjective estimates of temperature in England—to establish whether intervals reflected (and could be used to identify) this level of specificity. One-way repeated measures ANOVAs revealed significant effects of specifying time of year, upon both interval midpoints $F(2.079,81.093)=75.528$, $p<.001$, $\eta^2_p=.659$, and interval widths $F(2.260,88.141)=18.015$, $p<.001$, $\eta^2_p=.316$. Specifically, results found significantly higher absolute temperature estimates for July, and lower for December, as well as significantly lower interval widths for both of these months, by contrast with two questions (asked before and after the month specific questions) that did not specify a time of year.

Second, we examined whether intervals could discriminate a question with ambiguously interpretable phrasing from an unambiguous question, in this case relating to hypothetical increases in percentage recycling rates. Bootstrapped paired samples *t*-tests—used due to violations of normality—revealed significantly higher average interval midpoints when respondents were asked to *'increase* [10%] *by 50%'* than when asked to *'double'* this same initial rate, $p$=.016. However, despite mean interval width being approximately three times the size for the former question (.33 vs .11), this difference was not found to be statistically significant $p$=.161. These results are consistent with a bimodal distribution for the percentage question, comprising groups of respondents who interpreted *'increase by 50%'* differently—i.e., some interpreted it as a 50 percentage point increase, leading to a higher average response midpoint than the *'double'* question. The results do not however provide evidence that individual respondents commonly identified and reported the potential ambiguity in the question. These interpretations are supported by descriptive results showing the actual interval response distributions (see Fig. 9).

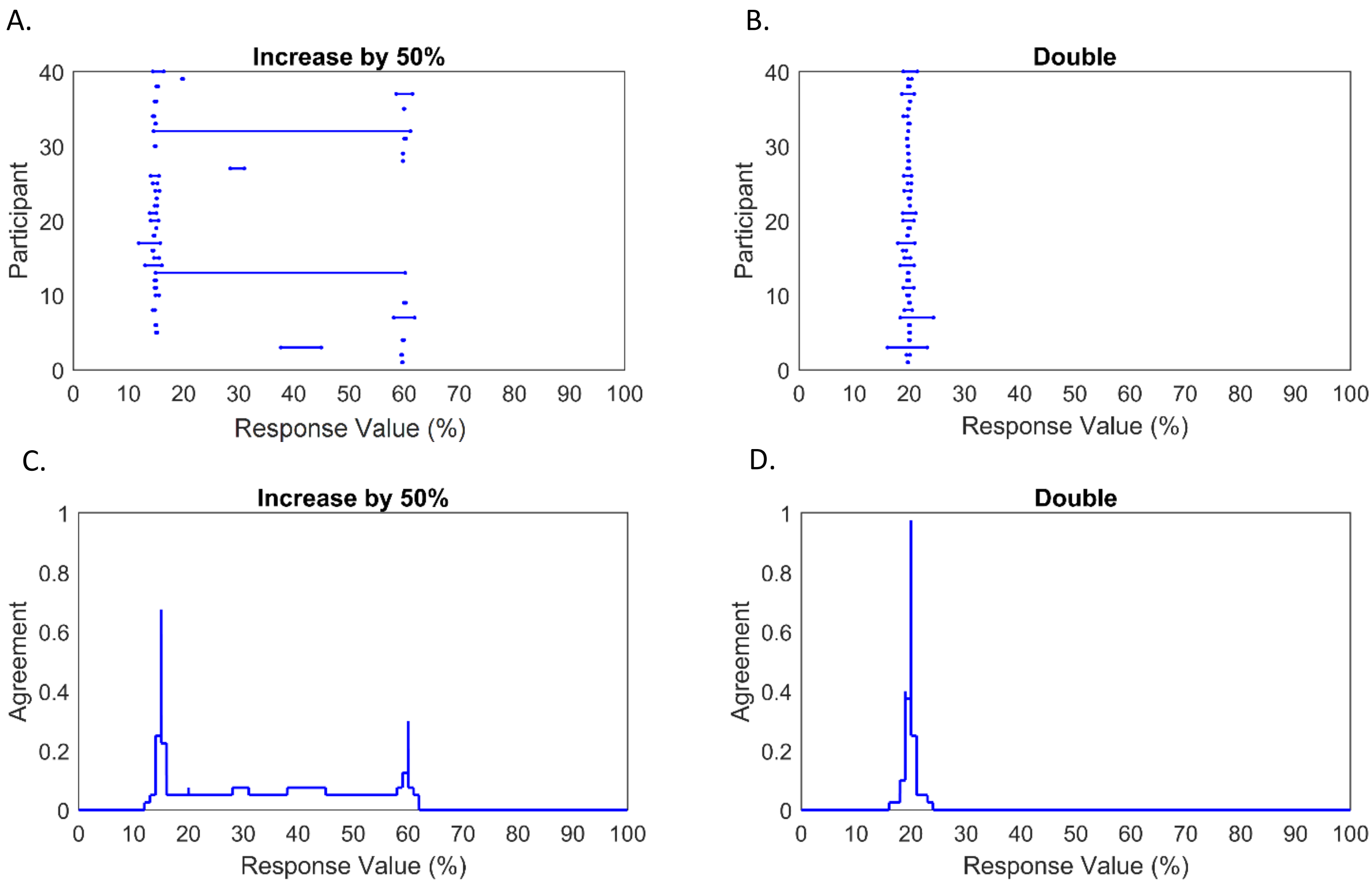

Figure 9: Responses to questions regarding hypothetical increases in recycling rates (beginning at 10%). Panels A & B—showing all interval-valued responses. Panels C & D—IAA plots showing aggregated interval-valued responses (rounded to nearest response integer).

Third, we examined whether intervals could identify questions within which a word was not understood by the respondent—likely to be of particular concern when designing surveys aimed at non-native speakers—in this case items related to personality characteristics. Two separate 4x3 repeated measures factorial ANOVAs were conducted, with dependent variables of interval midpoint and width, respectively. Independent variables were *characteristic* (four-levels: talkative, aggressive, lazy, quiet) and *word frequency* (three-levels: high, low, and pseudo-word—see *Questionnaire Items* section for more detail)—note that the four pseudo-words were randomly allocated between characteristic conditions. For interval midpoints, significant main effects were found for both characteristic $F(2.386,93.056)=5.250$, $p=.004$, $\eta^2_p=.119$, and word frequency $F(1.684,65.686)=7.352$, $p=.002$, $\eta^2_p=.159$, as well as a significant two-way interaction term

*F*(3.466,135.180)=8.900, *p*<.001, $\eta^2_p$=.186. For interval widths, there was a significant main effect of word frequency *F*(1.556,60.680)=59.922, *p*<.001, $\eta^2_p$=.606, but neither a significant main effect of characteristic *F*<1.0, nor a significant two-way interaction *F*(4.332,168.941)=1.166, *p*=.328. These findings reflect that respondents tended to identify with certain characteristics more than others. Also, on average, they were more likely to disagree with real words, but this was not the case for all characteristics—e.g., *'How well does the* [high frequency] *word talkative describe you?'* received more agreement than both its low frequency counterpart *'garrulous'* and control pseudo-word *'brendacious'*. Regarding interval widths, these were largest for pseudo-words, becoming progressively smaller for real but low frequency and then high frequency words—in a pattern highly consistent with expected levels of word comprehension.

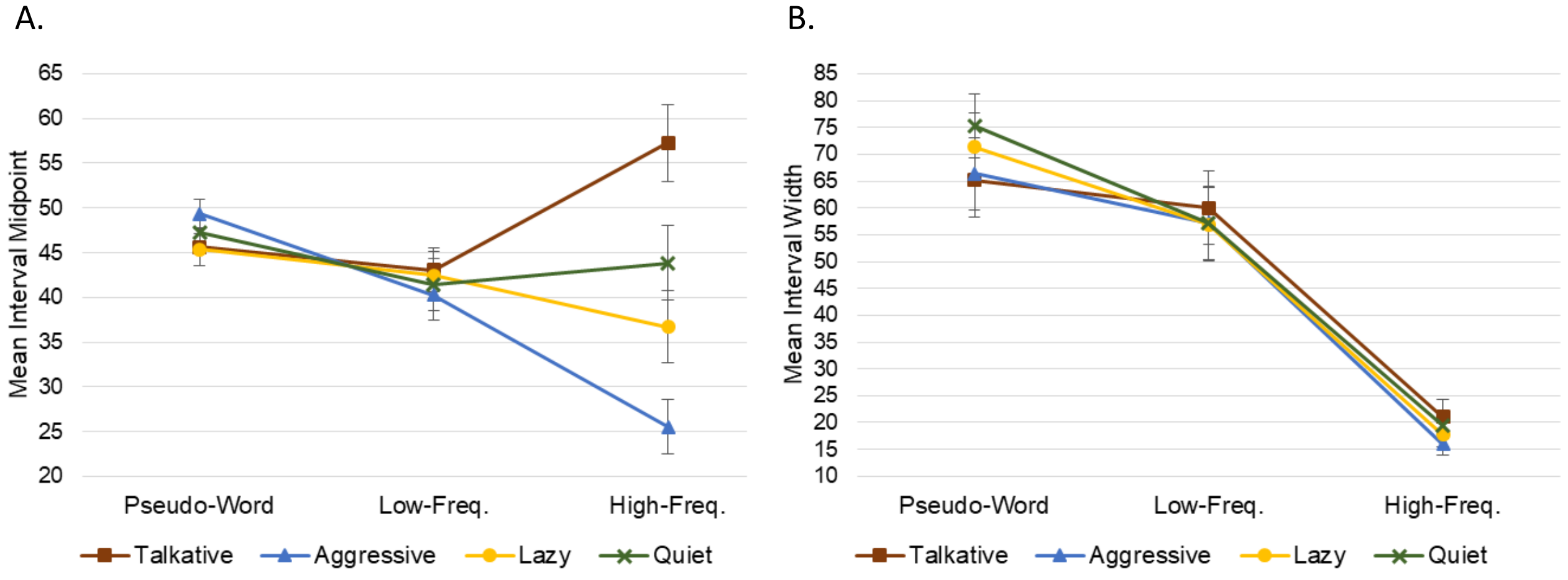

Figure 10: Showing mean interval midpoints (Panel A) and widths (Panel B) for each of the twelve personality questions. Error bars show 95% CIs.

Fourth, we used double- vs single-barrelled questions to examine in more depth whether intervals could discriminate between questions phrased precisely or more ambiguously. The same approach was taken as previously; two separate 4x3 repeated measures factorial ANOVAs were conducted, with dependent variables of interval midpoints and widths respectively. Independent variables were the *general topic* of the question (four-levels: reading books and watching television vs watching and playing sport vs drinking tea and coffee vs cooking and eating) and the *specific subject* of the question (three-levels: e.g., cooking and eating vs cooking vs eating). For interval midpoints, this revealed a significant main effect of topic $F(2.166,84.474)=15.483$, $p<.001$, $\eta^2_p=.284$, but no significant main effect of subject $F<1.0$ and no significant two-way interaction $F(3.254,126.887)=1.011$, $p=.395$. By contrast, for interval widths, this revealed a significant main effect of subject $F(1.428,55.687)=14.375$, $p<.001$, $\eta^2_p=.269$, but no significant main effect of topic $F(2.411,94.022)=2.257$, $p=.100$, nor a significant two-way interaction $F(<1.0)$. These results reflect significant differences in overall liking between topics, but not between the specific subjects within each topic. Nonetheless, significantly broader intervals were provided for more ambiguous double-barrelled items than for their single-barrelled counterparts.

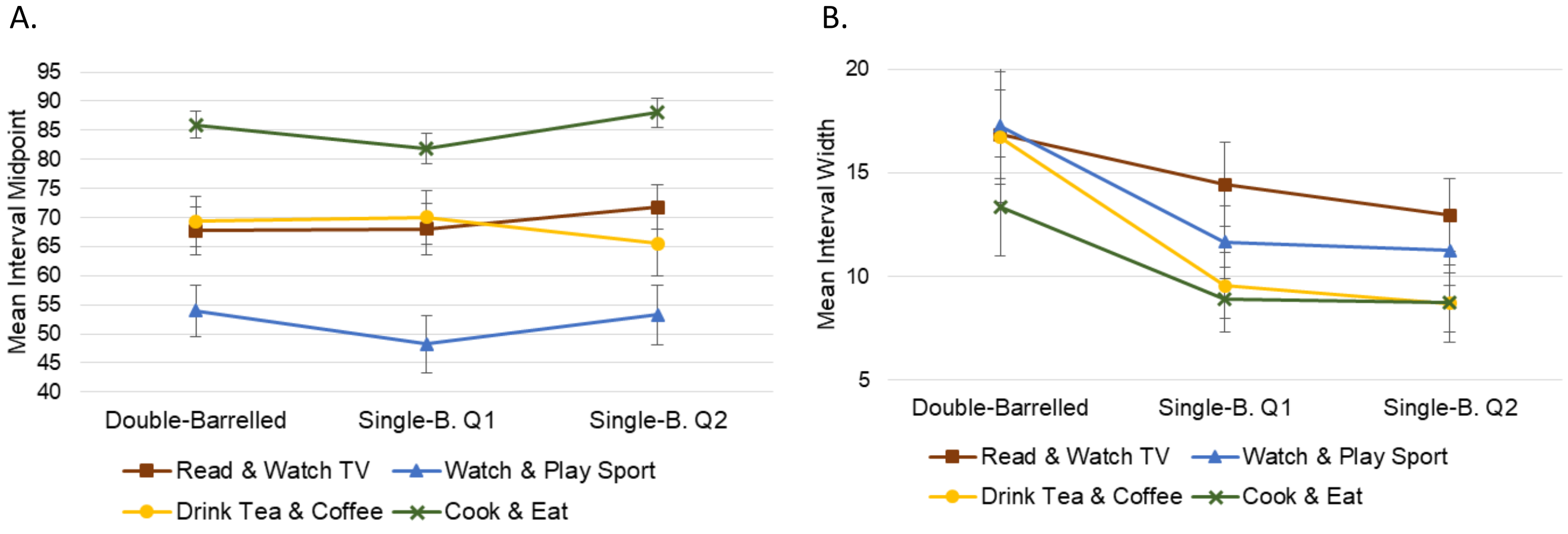

Figure 11: Showing mean interval midpoints (Panel A) and widths (Panel B) for each of the twelve double- vs single-barrelled questions. Error bars show 95% CIs.

Fifth, we examined judgements regarding the definition of stimuli falling within more or less ambiguous categories—in this case images of sets of eight vehicles, of which a potentially disputable number could be classified as 'cars' (see Fig. 5, both panels). Two further 2x2 repeated measures ANOVAs were conducted, with respective dependent variables of interval midpoints and widths. In this case, independent variables were the specific image stimulus used (i.e., Fig. 5 Panel A vs B), and the clarity of the category over which a judgement was made (i.e., cars vs vehicles). For interval midpoints, significant within-subject main effects were found for both image stimulus $F(1,39)=99.194$, $p<.001$, $\eta^2_p=.718$, and clarity of category $F(1,39)=85.403$, $p<.001$, $\eta^2_p=.687$, as well as a significant two-way interaction $F(1,39)=99.545$, $p<.001$, $\eta^2_p=.719$. These indicate that a greater number of stimuli were judged to be vehicles than cars in each image. Also, that a greater number of stimuli were judged to be cars in the second image (Fig. 5, Panel B), but an equal number were judged to be vehicles in each image. For interval widths, a significant main effect of category clarity was found $F(1,39)=7.374$, $p=.010$, $\eta^2_p=.159$, but neither a significant main effect of image stimulus $F(1,39)=1.746$, $p=.194$, nor a significant interaction term $F(<1.0)$. These reflect that broader intervals were provided for the relatively ambiguous category 'cars', across both image conditions, than for the more clearly inclusive category 'vehicles'.

Sixth, and finally, we examined perceptual judgements over more or less ambiguous image stimuli in another context—coloured marbles (see Fig. 6, all panels). Two one-way ANOVAs were conducted, with dependent variables of interval midpoint and width respectively. For interval midpoints, a significant effect of stimulus was found $F(1.498,58.418)=12.889$, $p<.001$, $\eta^2_p=.248$. The highest interval midpoints were provided for the colour-gradient set of marbles (Panel B: M=54.74), followed by the solid colour set (Panel A: M=50.21), and finally the patterned set (Panel C: M=42.05). Post-hoc paired-samples *t*-tests were conducted to examine paired differences, these were bootstrapped (10,000 samples) to ensure robustness to violations of normality. These revealed both increments to be statistically significant: $p=.011$, and $p=.005$.

For interval widths, a significant effect of stimulus was also found $F(1.490,58.126)=14.196$, $p<.001$, $\eta^2_p=.267$. The widest intervals were drawn for the patterned marble set (Panel C: M=20.56), followed by the colour-gradient set (Panel B: M=10.29), and the smallest intervals for the unambiguous, solid colours set (Panel A: M=1.90). Again, bootstrapped post-hoc paired-samples *t*-tests (10,000 samples) revealed each increment to be statistically significant: $p=.014$, and $p=.010$. This finding indicates that the width of participants' response intervals reflected the degree of uncertainty induced by each stimulus, as hypothesised. Results, illustrated using the IAA method, are shown in Figure 12.

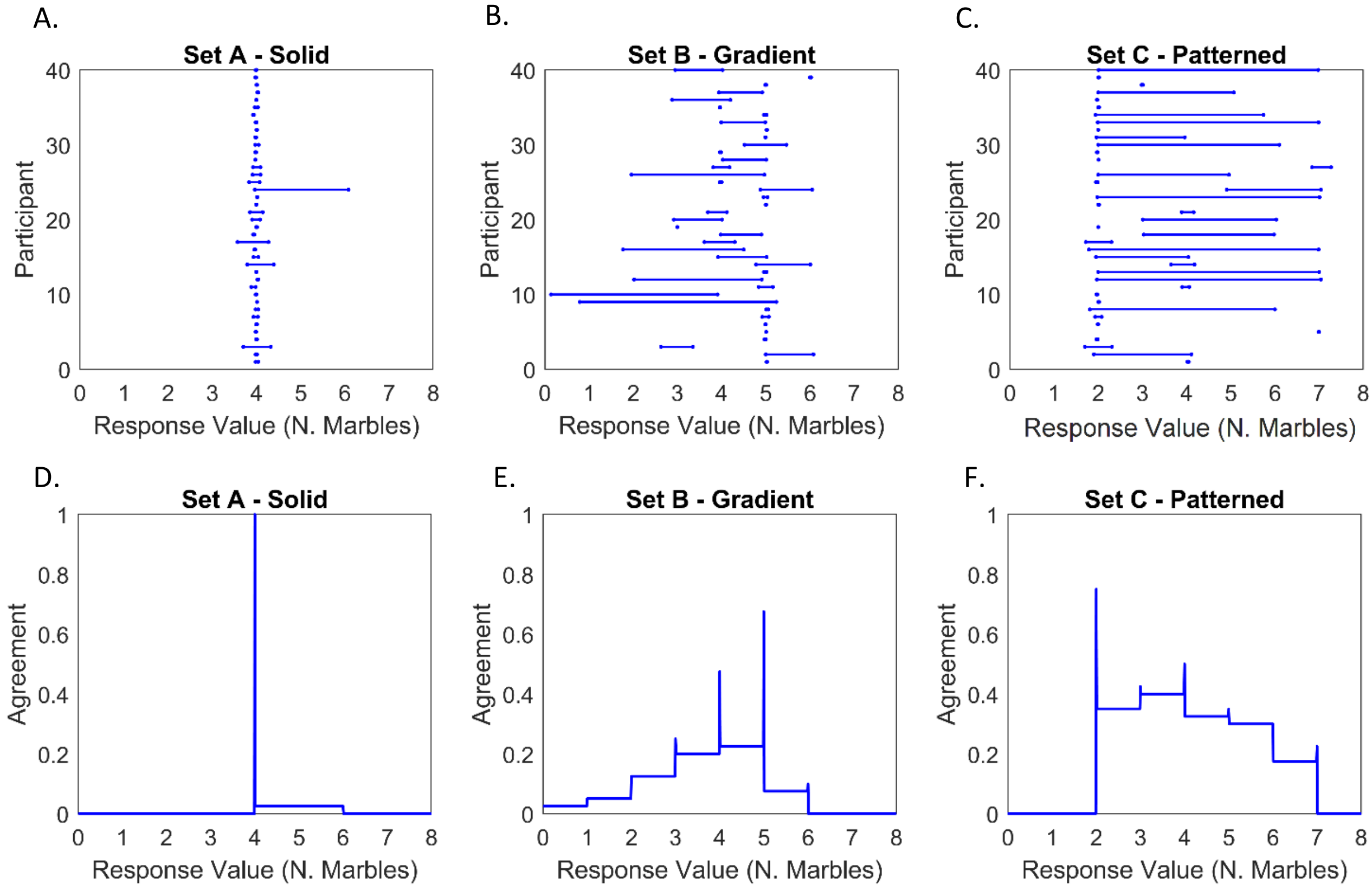

Figure 12: Responses to questions concerning the number of blue marbles contained in stimuli shown in Figure 6. Panels A, B & C—showing all interval-valued responses. Panels D, E & F—IAA plots showing aggregated interval-valued responses (rounded to nearest response integer).

**Subjective feedback:**

After finishing the main questionnaire, participants completed four subjective feedback questions, designed to assess their perceptions of the interval-valued response format. These were obtained using a traditional 5-point ordinal response scale (ranging from 1—Strongly Disagree, to 5—Strongly Agree). Descriptive results are shown in Table 3. By comparison with the midpoint of the response scale (3.00), these indicate significant agreement with the ellipse response-mode being easy to use, allowing effective communication of desired responses, and being liked overall, but significant disagreement with it being unnecessarily complex.

Table 3: Showing user feedback regarding the ellipse-based interval-valued response format.

| Question | Mean | SE |
|---|---|---|
| Easy to Use | 4.18 | .13 |
| Unnecessarily Complex | 2.05 | .14 |
| Effectively Communicate | 4.35 | .15 |
| Overall Liking | 4.53 | .10 |

## Discussion

The present study applied empirical methods to test the efficacy of a novel interval-valued survey response mode, in terms of capturing uncertainty and range in individual responses, arising from multiple sources. Capturing these is unfeasible using conventional discrete or point-valued response formats, such as Likert-type and Visual Analogue Scales, without asking specific follow-up questions, at the cost of substantially increased questionnaire duration and complexity. By contrast, we hypothesised that the proposed interval-valued response mode—which can now be administered digitally with a high degree of efficiency—can achieve this effectively, within a single coherent response, by providing participants with an additional dimension to their answers. Specifically, this response format enables respondents to vary the width of an interval, in the form of an ellipse drawn along a continuous scale, according to how certain and specific they wish to indicate each answer to be.

The core aim of this study was to determine whether the proposed interval-valued response mode is capable of systematically capturing vagueness and uncertainties in individual responses. This was done by examining interval-valued responses to questions of a specially designed survey, within which respondent knowledge (e.g., information availability), question clarity, and inherent variability in the correct response were each experimentally manipulated. In operational terms, it was hypothesised that variance in respondents' interval-widths would reliably reflect induced between-item variance in each of these factors.

Results consistently indicated in the affirmative, across a variety of contexts. In the first section of the survey, respondents used interval widths to effectively replicate observed differences in ranges of expected lifespan between animal species—accurately representing both facets of the data presented to them in interval format. This demonstrates the practical utility of this response mode as a quick and coherent method enabling participants to provide answers that inherently comprise a range (i.e., conjunctive sets). It also suggests that the response mode enables individuals to respond reliably in this context, even though this was not explicitly tested (e.g., through a test-retest design), and will require future study.

In the second section of the survey, respondents' intervals were found to consistently reflect their degree of uncertainty about a set of stimuli, in terms of information hidden from them, as well as inherent variability observed between cases within a given set. We note that the former here would result in a disjunctive set, representing the (lack of) knowledge of a true value, while the latter results in a conjunctive set describing an actual range of values (as no single discrete answer is correct). In turn, it is interesting to note that while each of these factors significantly contributed to response widths, results indicated that participants tended to substantially underestimate the former in their response intervals, mirroring a large body of evidence for overconfidence in probabilistic interval estimates (cf. Alpert & Raiffa, 1982; Juslin et al., 1999; Klayman, 1999; Soll & Klayman, 2004; Speirs-Bridge et al., 2010; Yaniv & Foster, 1995; 1997). By contrast, we found no such evidence for underestimation of conjunctive response ranges—which appeared well-calibrated. That is, while intervals provided did not tend to encompass all *possible* outcomes when a proportion of the set was hidden, they did closely represent the full range of *observed* outcomes within a set—at least when a high proportion of marbles were visible. This effect was evident in certain cases as an otherwise incongruous increase in mean interval widths corresponding with an increase in available information, as greater disparity was revealed within a set than may have been expected (see Fig. 8, Sets 4 & 6). Future research might be able to leverage a conjunctive set framing as a pathway to reducing overconfidence.

It was also found that when there was a high amount of both hidden information and stimulus variability this led to further underestimation of total response width; this may reflect participants' difficulty in dealing with these two sources of interval width together (i.e., forcing them to effectively integrate disjunctive and conjunctive sets) exacerbating the existing tendency to underestimate epistemic uncertainty. Alternatively, it may reflect a deliberate interaction between the two factors, whereby lower information about the set of stimuli leads to lower confidence in the reliability of any variability that is currently observed—that is, participants may have assumed that observed discrepancies were more likely to even out once more hidden marbles were revealed than they were to increase.

On the whole, these findings demonstrate the capacity of the proposed response format to efficiently capture and quantify degrees of both disjunctive and conjunctive range in responses—although it is not designed to discriminate the two. This could prove valuable across a wide range of circumstances where assessments of real-world variables, solicited from experts or the broader population, are crucial to making predictions and informing subsequent policy, strategy, or investment decisions—e.g., political polling, supplier selection, marketing, environmental planning, cyber-security (i.e., vulnerability) and other risk or impact assessments. Likewise, in the context of social, behavioural, and psychological research, intervals may facilitate richer capture and understanding of participants' preferences, attitudes, or choices, considering the relevant information available to them, and in combination with more subjective factors such as beliefs and their associated confidence or conviction. Intervals also provide the necessary information to guide potential follow-on qualitative research stages, such as to explore the underlying origins and drivers of uncertainties or ambiguities once they have been identified.

In the third section of the survey, respondents were found to systematically vary their interval sizes across a broader range of situations—each relating to their degree of subjective uncertainty, induced by the wording of the question or its associated stimulus. Participants drew significantly smaller intervals to represent temperature estimates for specific months than for non-time-specific questions. But while respondents were inconsistent in their interpretations of a deliberately ambiguous question ('*increase* [10%] *by 50%*'), only a minority explicitly recognised and communicated the potential ambiguity in their response by providing a range from 15% to 60%. When presented with personality judgements concerning adjectives of varying degrees of obscurity, interval widths reflected word frequency in the manner expected. For high frequency words interval widths fluctuated between respondents and questions, but these remained relatively small compared with the other conditions, with no respondents circling the entire scale. For low frequency words mean widths were substantially greater, indicating uncertainty, with approximately half of responses covering the entire response-scale, indicating that the appropriate response was unknown. For fabricated pseudo-words, this method of indicating complete uncertainty became the majority response—although it is worth noting that some respondents did select only the centre-point in these cases. It is likely that these participants reverted to a habitualised response, as would be given on a discrete response-scale. This may be a greater problem over longer questionnaires, for which response fatigue may increase this

effect—but it can be expected that, as the interval-valued response mode becomes more widely used, familiarity with this format will overcome any inertia working against its appropriate use.

In addition to this, interval widths for double-barrelled questions were also found to be significantly broader than those for their more specific single-barrelled counterparts. Furthermore, questions requiring judgement over stimulus category membership found significantly broader intervals for a contextually ambiguous category term (*car*) than for a more clearly inclusive term (*vehicle*). Likewise, significantly larger intervals were provided for membership of the same category term (*blue*) when the stimuli over which the judgement was made were less clearly classifiable. These cases highlight the potential for interval-valued responses to identify questions which respondents struggle to understand—whether this be due to unclear stimuli or imprecise question wording. This capability could be of great value, for example in questionnaire pre-testing, particularly in cases where a questionnaire will be taken by non-native speakers.

Participants' subjective feedback was also collected and assessed. This was found to be generally positive on all four questions asked. On average, respondents agreed both that the response format was easy to use and that it allowed them to effectively communicate their desired response, while they disagreed that it was unnecessarily complex. Overall liking received greater agreement than any of the more specific questions, which may be due to some respondents having been more uncertain in their responses to the sub-components, as a result of their greater complexity. Ironically, capturing interval-valued feedback, rather than using a traditional discrete response scale, may have shed more light on this effect.

**Summary and Future Work**

To summarise, in the present paper we put forward and empirically examine a new ellipse-based questionnaire response format. This is designed to capture interval-valued data with a focus on efficiency and user experience, including through increasingly ubiquitous information technologies, minimising the time and effort requirements for both survey respondents and administrators. First, we present a case for the potential added value of this response mode, by contrast with conventional alternatives. Here, we identify three primary situations in which interval-valued responses might capture new and valuable information: where responses inherently comprise a range of values, where they are uncertain due to a lack of knowledge about the correct answer, and where they are ambiguous, or otherwise uncertain, due to imprecision or ill-definition in either the question or associated stimulus.

Following this, we document a validation study within which interval-valued responses are collected across a range of questionnaire items—designed to exemplify each of the three cases described above. Interval widths were found to systematically reflect experimentally induced differences between items, in relation to each of these factors. These findings provide a basis of empirical evidence for the efficacy and practical value of this new interval-valued response mode, but further research is needed to determine the extent to which these generalise—assessing the impact of individual, or sample, differences on the realisation of this value in practice; as well as exploring the effects of variations in interval-valued scale design upon outcomes including data fidelity, reliability, and participant workload.

Future work is also required to more thoroughly explore the broad range of potential benefits offered by this information-rich type of response—as well as to establish how to best retain and interpret this additional information, through development of novel methods of statistical analysis, which may build upon a broad body of existing work across academic disciplines including symbolic data analysis, interval arithmetic, interval ranking, fuzzy sets, and interval aggregation through fuzzy integrals (cf. Anderson et al., 2014; Billard, 2006; Billard & Diday, 2003; Ferson, Kreinovich, Hajagos, Oberkampf & Ginzburg, 2007; Liu & Mendel, 2008; McCulloch et al., 2019; Miller et al., 2012; Nguyen et al., 2012; Wagner et al., 2013; Wagner et al., 2015; Wu et al., 2012). Another practical concern remains in demonstrating the efficacy of this approach across a broader variety of real-world contexts—expanding upon a growing body of research (cf. Ellerby, McCulloch, Wilson & Wagner, 2019; Ellerby et al., 2020; Navarro et al., 2016; Wallace, Wagner & Smith, 2016). Finally, it is important that future empirical research is conducted to directly compare this interval-valued response mode against existing methods of both lower (e.g., Likert, VAS) and higher complexity (e.g., multi-step interval elicitation, SHELF, FRS)—to establish and quantify the trade-offs between information capture and important practical concerns regarding response efficiency (i.e., workload, survey duration, perceived complexity, training requirements), which have a large influence on likelihood of broader adoption.

**Acknowledgement of Support:**

This research was supported by the UK's Engineering and Physical Sciences Research Council (EPSRC) research grant EP/P011918/1.

**Open Practices Statement:**

The DECSYS software used to implement the survey documented in this paper is accessible via https://www.lucidresearch.org/decsys.html The software is hosted on GitHub to invite community engagement and contributions under the open-source GNU Affero licence. Full code and documentation are stored at the following repository https://github.com/decsys/decsys

Interval-valued data and commented code for extraction and creation of IAA charts and interval plots—as shown in Figs. 7, 9 & 12—are stored at https://osf.io/pb4ne/ to provide a working example of this process. Original survey stimuli are also available via the same link.

Further original study materials and data are available from the authors upon reasonable request, and with commitment to abide by applicable data protection regulations. The experiment was not preregistered.

**Footnotes:**

[1] Uncertainty can be broken down into epistemic (reducible, subjective) and aleatoric (irreducible, objective), (cf. Brugnach, Dewulf, Pahl-Wostl, & Taillieu, 2007; Couso & Dubois, 2014; Dewulf, Craps, Bouwen, Taillieu, & Pahl-Wostl, 2005; Dubois & Prade, 2012; Fox & Ülkümen, 2011; Kwakkel, Walker & Marchau, 2010; Walker et al., 2003; Zandvoort, Van der Vlist, Klijn, & Van den Brink, 2018). The second dice example is of aleatoric uncertainty, as the correct response is yet to be determined and is therefore currently unknowable. For an example of epistemic uncertainty, relating instead to a lack of knowledge regarding a knowable answer, consider the question 'I have just rolled two fair six-sided dice one time, what sum total did I roll?' when the dice remain hidden from the respondent's view. Here the answer is knowable but remains uncertain from the perspective of the respondent, until more information can be obtained (e.g., viewing the dice in question). For the purposes of the present study, we focus on the breadth of underpinning 'drivers' of interval width, which may be encountered when completing a conventional questionnaire, rather than arguing for a particular typology of uncertainties. It is thus convenient to note that both of these cases fall within the single category of disjunctive sets—i.e., the response interval reflects the inadequacy of knowledge or information necessary for the respondent to make a precise and certain judgement.

[2] Here the term interval refers to the definition designated by Stevens (1946). This is semantically different from how this term is more commonly used throughout the rest of this paper, and indeed most of the sciences. Outside of this section an 'interval' will refer to the mathematical notion of a closed set of values on a continuous real line, which is defined by two endpoints and possesses the property of width, which is zero only in the case that both endpoints are identical.

[3] Note that as these stimuli were shown on an LED computer screen, a perceiver with high enough image resolution could argue that none of the marbles truly contained any yellow—as this percept is created through the combination at a specific ratio of green and red light output.

[4] Participants were instructed to judge the number of blue, rather than green or yellow marbles because of the relatively high prevalence of red-green colour deficiency (up to 8% in European males—Birch, 2012) relative to blue-yellow (<0.01% worldwide—Wright, 1952).

[5] This software is available as an open-source package which enables interested parties to run their own instance of the platform to conduct surveys. Beyond this, DECSYS is designed to enable hosting for example on standard cloud infrastructure such as Microsoft Azure, providing a similar user experience to commercial tools such as SurveyMonkey® or Qualtrics®. Here, the software provides the functionality of multiple survey-administrator accounts—allowing, for example, individual researchers to develop and administer their own surveys and access the resulting data. All data here are accessible exclusively by the specific survey-administrator. In a geographical sense, the physical location of the data stored is determined by whether DECSYS is being used as a service hosted on a personal server (e.g., within a university), or on cloud infrastructure. In the case of the present study, all data were stored locally.

[6] This enables analysis at a reasonable level of complexity, using established methods. However, in general, it is complete intervals which capture individual responses. With the application of a rapidly advancing body of work around advanced techniques for handling interval-valued data we expect that there is significant potential in the development of new interval-based quantitative analysis methods, offering an efficient approach to rapidly collate, compare, and assess rich quantitative information from survey respondents.

[7] The interval midpoint is used to represent interval location in our analysis, and we do not assume that the midpoint of each interval is necessarily a 'best estimate'. In cases where the response interval represents a disjunctive set, a respondent's 'best estimate' may or may not be located at the centre of the interval provided.

## Appendix A: Instructions for the Ellipse Response Mode

Thank you for agreeing to take part in this experiment.

This task is designed to assess how well people are able to respond using a newly developed survey response-format. Here, responses are made by drawing ellipses on a continuous scale, instead of picking a single response option. How to use this type of response-format is explained below, with examples. Please read this information carefully and ask the experimenter if you have any questions.

**For each question:**

- Use an ellipse to mark your answer, surrounding the area of the scale that you wish to select. A smaller or wider ellipse can be used to indicate the level of uncertainty, range or vagueness in the response.
- For instance, if you feel unable to precisely answer the question quickly and satisfactorily with the information provided, then use a larger ellipse to represent the uncertainty in your answer.
- Even if you have all of the information that you need to give an accurate response, the most appropriate answer may include a range of values. In this case, you should draw an ellipse that includes all values that represent a correct response.
- If you do not know the answer at all, then you should circle the entire scale.
- If you wish to indicate a certain and completely precise answer, then you can do so by drawing a single vertical line.

1. An uncertain response

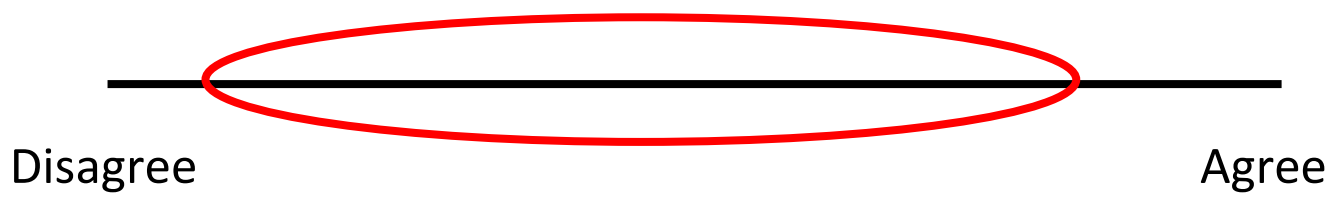

2. A more certain response

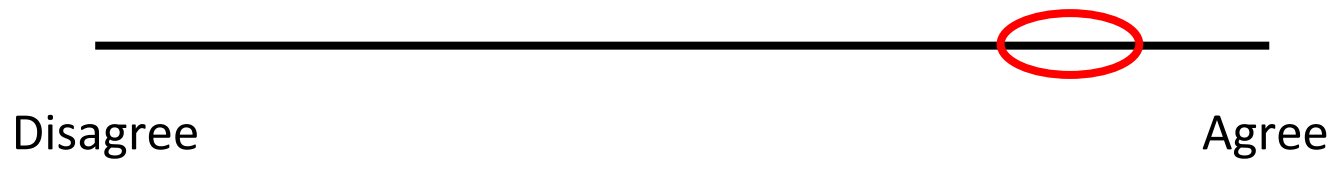

3. A don't know response

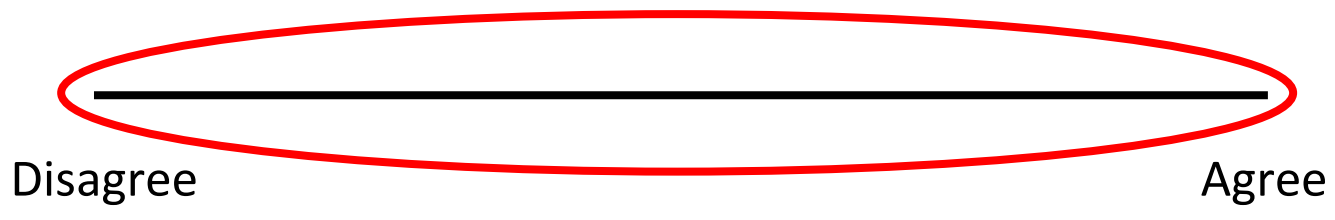

4. A completely certain response

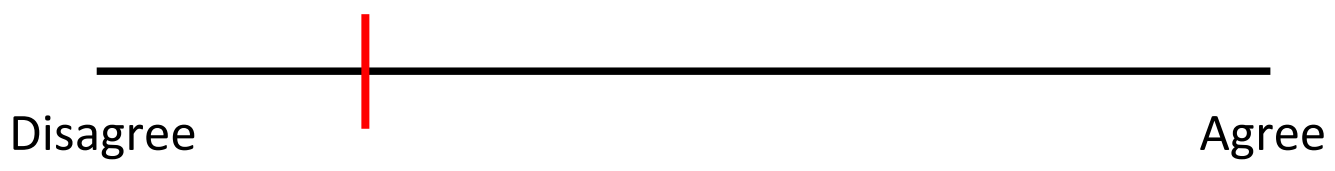

**Appendix B: Survey Instructions (text content)**

There will be three sections to this survey, each with slightly different tasks. Examples of how you can respond are provided for each.

Section 1: Here you will be provided with a chart, showing the life expectancies of different animals. In a series of questions, you will be asked to correctly identify the life expectancy of an animal from this chart, and to communicate this as accurately as you can, through drawing an ellipse on the scale provided.

Section 2: Here, you will be shown a series of sets of marbles. Each set will be made up of 5 rows of 7 marbles each, but some may be hidden. All marbles will be either blue or yellow and you will be asked to estimate the number of blue marbles that are in each row. You will only make one estimate per set of marbles (5 rows). If any marbles are hidden then you should make your estimate including these hidden marbles. You can illustrate your degree of uncertainty using the width of the ellipse.

Section 3: Here, you will be asked a series of more varied questions. Respond as you feel appropriate, applying the same approach as in the previous sections.

Please indicate when you have read and fully understood this information sheet, and you are ready to proceed. Ask the experimenter if you have any further questions.

**Appendix C: Monte-Carlo ANOVA results (robustness checks)**

*Time of year, one-way (4-level) ANOVAs*

Midpoints $p<.001$

Widths $p<.001$

*Personality words, 4x3 ANOVAs*

Midpoints: Characteristic $p=.002$, Word Freq. $p<.001$, Int. $p<.001$

Widths: Characteristic $p=.638$, Word Freq. $p<.001$, Int. $p=.318$

*Double-barrelled questions, 4x3 ANOVAs*

Midpoints: Topic $p<.001$, Subject $p=.398$, Int. $p=.414$

Widths: Topic $p=.082$, Subject $p<.001$, Int. $p= .632$

*Vehicle questions, 2x2 ANOVAs*

Midpoints: Stimulus $p<.001$, Category Clarity $p<.001$, Int. $p<.001$

Widths: Stimulus $p=.190$, Category Clarity $p=.004$, Int. $p=.359$

*Marble questions, one-way (3-level) ANOVAs*

Midpoints $p<.001$

Widths $p<.001$